\documentclass[12pt]{article}

\usepackage{epsfig}
\usepackage{epstopdf}
\usepackage{amssymb,amsmath}
\usepackage{comment}
\usepackage{enumerate}
\usepackage{multirow}
\usepackage{multicol}

\usepackage[sorting=none,citestyle=numeric-comp]{biblatex}
\usepackage{hyperref}

\newcommand{\bea}{\begin{eqnarray}}
\newcommand{\eea}{\end{eqnarray}}

\numberwithin{equation}{section}

\begin{document}

\begin{titlepage}
%\begin{flushright}
%
%\end{flushright}
%
\vspace*{10mm}
\begin{center}
\baselineskip 25pt 
{\Large\bf
%%%%%%%%%%%%%%%%%%%%%%%%%%%%%%%%%%%%%%%%%%%%%%%%%%%
Analytic formulation of Leptogenesis with
neutrino oscillation data employing the general 
parametrization for neutrino mass matrix
%%%%%%%%%%%%%%%%%%%%%%%%%%%%%%%%%%%%%%%%%%%%%%%%%%%
}
\end{center}
\vspace{5mm}
\begin{center}
{\large
Nobuchika Okada$^{~b,}$\footnote{okadan@ua.edu}, 
and Digesh Raut$^{~e,}$\footnote{draut@smcm.edu}
}
\end{center}
\vspace{2mm}

\begin{center}
{\it
$^{a}$ Department of Physics and Astronomy, \\ The University of Alabama, Tuscaloosa, Alabama 35487, USA \\
$^{b}$ Physics Department, \\St. Mary's College of Maryland, 
St. Mary's City, Maryland 20686, USA
}
\end{center}
\vspace{0.5cm}
%%%%%%%%%%%%%%%%%%%%%%
\begin{abstract}
%%%%%%%%%%%%%%%%%%%%%%
The observed neutrino oscillations and baryon asymmetry, unexplained by the Standard Model (SM), can both be accounted for by extending the SM to include Majorana right-handed neutrinos (RHNs). Tiny neutrino masses naturally arise through the Type-I seesaw mechanism, which involves lepton number violation. Meanwhile, the baryon asymmetry can be generated via leptogenesis, where the out-of-equilibrium decay of RHNs produces a lepton asymmetry that is partially converted into a baryon asymmetry through sphaleron processes. The Dirac Yukawa couplings play the crucial role for both Type-I seesaw and leptogenesis. 
In this work, we derive an analytic expression for the CP asymmetry parameter in a general parametrization. Focusing on a hierarchical RHN mass spectrum, we evaluate the lowest mass of the lightest RHN that reproduce both neutrino oscillation data and the observed baryon asymmetry. We study the case with two and three generations of RHNs for both thermal and non-thermal leptogenesis scenarios. Besides the standard Type-I seesaw involving SM Higgs doublet, we also examine the Type-I seesaw with a new neutrinophillic Higgs doublet.  
In this case for non-thermal leptogenesis, the minimum value for the lightest RHN mass can be as low as sphaleron decoupling temperature. 

%Since this coupling can be enhanced via a general parametrization, the predicted mass of the lightest RHN can potentially be lowered. 
   
%%%%%%%%%%%%%%%%%%%%%%%
\end{abstract}
\end{titlepage}

%%%%%%%%%%%%%%%%%%%%%%%%%%%%%%%%%%%%%%%%%%%%%%%%%%%%%%%%%%%
\section{Introduction}
%%%%%%%%%%%%%%%%%%%%%%%%%%%%%%%%%%%%%%%%%%%%%%%%%%%%%%%%%%%
The observation of neutrino oscillations \cite{pdg2024} have established that neutrinos have tiny but non-zero masses and flavor mixings, which cannot be explained within the framework of the Standard Model (SM). The Type-I seesaw mechanism \cite{Minkowski:1977sc,Yanagida1979,GRS1979,Glashow1979,Mohapatra:1979ia,Schechter:1980gr} provides the simplest scenario to naturally explain the origin of tiny neutrino masses by extending the SM with gauge-singlet Majorana right-handed neutrinos (RHNs). The observed light neutrino masses is given by the seesaw formula,   
\bea
m_\nu \simeq m_D \left(\frac{m_D}{ m_N}\right), 
\label{eq:seesaw1}
\eea
with $m_D \ll m_N$, where $m_D = Y_D v_h/\sqrt{2}$ is the Dirac type neutrino mass  generated by electroweak symmetry breaking and $m_N$ is the RHN mass.  
Cosmic microwave background measurements constraint $\Sigma_i \; m_{\nu^i} \leq 0.083$ eV \cite{pdg2024}.

RHNs can also provide an explanation for the origin of the observed matter–antimatter asymmetry in the universe, a long-standing puzzle that remains unresolved within the SM framework.
The mechanism is known as leptogenesis \cite{Fukugita:1986hr}. In leptogenesis, the out-of-equilibrium decay of RHNs, which violates both $C$ and $CP$ symmetries, generates a lepton asymmetry. Sphaleron process \cite{tHooft:1976rip, Manton:1983nd, Klinkhamer:1984di} converts this lepton asymmetry to baryon asymmetry \cite{Kuzmin:1985mm, Arnold:1987mh, Khlebnikov:1988sr, Harvey:1990qw},
\bea
\frac{n_B}{s} = -\left(\frac{28}{79}\right) \left(\frac{n_{L}}{s}\right),  
\label{eq:nB}
\eea
where $28/79$ is the sphaleron conversion factor, $n_B \equiv n_b - n_{\bar{b}}$ and $n_L \equiv n_\ell - n_{\bar{\ell}}$ are the baryon and lepton number densities, and $s$ is the entropy density of the universe. 
The observed baryon asymmetry as determined from cosmic microwave background (CMB) measurements is given by \cite{pdg2024} 
\bea
\frac{n_B}{s} =  8.7 \times  10^{-11},  
\label{eq:nBexp}
\eea
whereas the lepton asymmetry generated by RHN decay can be estimated as 
\bea
\frac{n_L}{s} \simeq  \frac{\epsilon}{g_*} \times \kappa\; .   
\label{eq:nL}
\eea
Here, $\epsilon$ is the CP-asymmetric parameter, $\kappa < 1$ quantifies the dilution of the lepton asymmetry due to inverse processes active during RHN decoupling, and $g_\star^{SM} \simeq 106.75$ is the effective number of relativistic degrees of freedom (d.o.f) in the thermal plasma at decoupling. 
Hence, to reproduce the observed baryon asymmetry, $\epsilon$ must satisfy
\bea
\epsilon \;  \simeq\;  \frac{1}{\kappa} \times 10^{-8}. 
\label{eq:epsbound}
\eea 

The CP asymmetry in RHN decay is generated by the interference between tree-level and one-loop level amplitudes and it's magnitude is determined by the strength of the Dirac Yukawa coupling \cite{Covi:1996wh, Buchmuller:1997yu}. 
It can be estimated as
\bea
\epsilon \;\simeq\; \frac{3}{16\pi}Y_D^2.  
\eea
Using the $\epsilon$ required to reproduce the observed baryon asymmetry, we obtain 
\bea
Y_D^2 \;\simeq\;10^{-7} \times \frac{1}{\kappa}. 
\label{eq:YD}
\eea
The Dirac Yukawa coupling can be expressed in terms of neutrino masses using the seesaw relation in Eq.~(\ref{eq:seesaw1})  
\bea
Y_D^2 \simeq \frac{2 m_\nu}{v_h^2}\times m_N.   
\eea
where $v_h = 246$ GeV is the vacuum expectation value (VEV) of the SM Higgs. 
Using the last two relations together with a conservative estimates for $m_\nu$ and $\kappa$, we obtain a lower bound on the $m_N$, 
\bea
m_N \gtrsim 10^{10} {\rm GeV} \times \left(\frac{0.05 {\rm eV}}{m_\nu}\right) \left(\frac{v_h}{246 \; {\rm GeV}}\right)^2 \left(\frac{0.01}{\kappa}\right). 
\eea
This result is consistent with the previous numerical studies, $m_N \gtrsim 10^{9}-10^{10}$ GeV \cite{Davidson:2002qv,Buchmuller:2004nz,Buccella:2012kc}

In the Type-I seesaw framework, the Dirac Yukawa coupling matrix $Y_D$ is defined up to an arbitrary complex orthogonal matrix \cite{Casas:2001sr, Ibarra:2003up}, which is independent of neutrino oscillation data. 
It may therefore be possible to enhance the Dirac Yukawa coupling $Y_D$ to lower the bound on $m_N \gtrsim 10^{9}-10^{10}$ GeV. 
In this work we derive an analytic expression for the CP asymmetry parameter $\epsilon$ by explicitly incorporating the orthogonal matrix, and re-examine the lower bound on $m_N$ by taking into account all of its free parameters. 
We will consider scenarios involving two and three generations of RHNs, examining both thermal and non-thermal leptogenesis frameworks. 
In addition to the Type-I seesaw where the RHN couples with SM Higgs and leptons, we will also consider the seesaw where the RHN couples with SM leptons and a new Higgs doublet, so-called leptophillic Higgs doublet,  instead of the SM Higgs doublet \cite{Ma:2000cc}. 
In this case the electroweak VEV $v_h$ in Eq.~(\ref{eq:YD}) is replaced with the VEV of the new Higgs doublet $v_2$ with $v_2 \gtrsim 10$ MeV to avoid observable lepton flavor violation \cite{Guo:2017ybk}.

The paper is organized as follows: In Sec.~2, we discuss the Type-I seesaw scenario with two and three generations of RHNs. Sec.~3 presents analytic expressions for the CP asymmetry parameter in both cases. In Sec.~4, we present results for thermal and non-thermal leptogenesis. Sec.~5 discusses the Type-I seesaw and Leptogeneis in a  neutrinophilic Higgs doublet model. Our result is summarized in Sec.~6.

%%%%%%%%%%%%%%%%%%%%%%%%%%%%%%%%%%%%%%%%%%%%%%%%%%%%%%%%%%%
\section{Type-I Seesaw}
%%%%%%%%%%%%%%%%%%%%%%%%%%%%%%%%%%%%%%%%%%%%%%%%%%%%%%%%%%%
%The type-I seesaw mechanism \cite{seesaw} naturally explains the origin of tiny neutrino masses by introducing heavy Majorana right-handed neutrinos (RHNs) as SM gauge singlets. The extended SM Lagrangian, 

To implement the Type-I seesaw, SM singlet right-handed Majorana neutrinos are added to the SM particle content. 
The SM Lagrangian is extended to include the following terms:    
\bea  
   {\cal L} \supset  \left(-\sum_{i = 1}^{3} \sum_{j} Y_D^{ij} 
   \overline{N_R^i} \; H^\dagger \ell_L^j   - \frac{1}{2} \sum_{j} m_{N^j}  \overline{N_{R}^{j~C}} N^j_{R} \right) +{\rm h.c.},
   \label{eq:seesawLag}
\eea 
where $Y_{D}^{ij}$ is the Dirac-type Yukawa involving $j$ generations of Majorana neutrinos ($N_R^j$) with mass $m_{N^j}$. 
%where the first and the second terms are, respectively, the Dirac and Majorana Yukawa couplings, and we have chosen the flavor diagonal basis for the Majorana Yukawa couplings. 
%For the rest of this work, we will set $Y_M^{1,2,3} \equiv Y_M$ for simplicity. 
%After the breaking of the electroweak and the $B-L$ gauge symmetries, the  Dirac and Majorana masses for the neutrinos are generated in Eq.~(\ref{eq:seesaw}):
After electroweak symmetry breaking the the vacuum expectation value (VEV) of SM Higgs, $\langle H\rangle = ( v_h/\sqrt{2}, 0) ^{T}$, we obtain 
\bea  
   {\cal L} \supset   \left(-\sum_{i = 1}^{3} \sum_{j} m_D^{ij} \overline{N_R^i} \; \nu_L^j  - \frac{1}{2} \sum_{j} m_{N^j} \overline{N_{R}^{j~C}} N^j_{R} \right) +{\rm h.c.},
\eea 
where $m_D^{ij}  = \frac{Y_D^{ij}}{\sqrt{2}} v_h$ and we have chosen then basis in which the RHN mass matrix is diagonal. 
The resulting neutrino mass matrix is given by 
\bea
M_\nu = \begin{pmatrix}
0 & m_D^T \\
m_D & M_N 
\end{pmatrix}, 
\eea
where $m_D$ and $M_N$ are Dirac and Majorana mass matrices, respectively. 
Assuming a hierarchy, $m_N^{j}\gg m_D^{ij}$, the neutrino mass matrix can be approximately block-diagonalized. 
The heavy neutrinos mass is approximately given  by $M_N$, while the light netutrino mass matrix is determined by the seesaw formulas 
\bea
m_\nu \simeq m_D^T M_N^{-1} m_D. 
\label{eq:seesaw}
\eea
To diagonalize $m_\nu$, we use the unitary Maki–Nakagawa–Sakata (MNS) matrix \cite{Maki:1962mu},  
\bea
D_\nu \equiv {\rm diag} (m_1, m_2, m_3) = U_{MNS}^T \; m_\nu \; U_{MNS},  
\label{eq:Dnu}
\eea
where $m_i$ are the observed masses of the light neutrinos. 
In the case of three RHNs, the explicit form of the MNS matrix is given by 
\bea
U_{\rm MNS}
=
\left(\begin{array}{ccc}c_{12}c_{13} & s_{12}c_{13} & s_{13}e^{-i\delta} \\-s_{12}c_{23}-c_{12}s_{23}s_{13}e^{i\delta} & c_{12}c_{23}-s_{12}s_{23}s_{13}e^{i\delta} & s_{23}c_{13} \\s_{12}s_{23}-c_{12}c_{23}s_{13}e^{i\delta} & -c_{12}s_{23}-s_{12}c_{23}s_{13}e^{i\delta} & c_{23}c_{13}\end{array}\right)
\left(\begin{array}{ccc}e^{i\sigma_1} & 0 & 0 \\0 & e^{i\sigma_2} & 0 \\0 & 0 & 1\end{array}\right), 
\eea
where
$s_{ij}=\sin\theta_{ij}$ and  $c_{ij}=\cos\theta_{ij}$ are mixing angles,
$\sigma_i$ ($\delta$) are the CP-violating Majorana (Dirac) phases.  
These are constrained by neutrino oscillations measurements. 
In our analysis we will use the result from a global fit presented in Ref.~\cite{Esteban:2024eli}: 
\bea
&& \sin^2 \theta_{12}= 0.307, \quad 
\sin^2 \theta_{13}= 2.20 \times 10^{-2}, \quad 
\Delta m_{21}^2 = 7.53 \times 10^{-5}\;  {\rm eV}, \quad 
\delta = 1.23 \pi \; {\rm rad},  \nonumber \\
&& {\rm (NH)} \sin^2 \theta_{23}= 0.534, \quad  
\; {\rm (NH)} \Delta m_{32}^2= 2.437\times 10^{-3} \;  {\rm eV^2}, \nonumber \\
&& {\rm (IH)} \; \sin^2 \theta_{23}= 0.547, \; \quad \; {\rm (IH)} \; \Delta m_{32}^2= -2.519\times 10^{-3} \;  {\rm eV^2} \qquad, 
\label{eqn:oscidata}
\eea
where $\Delta m_{ij}^2 = m_i^2 - m_j^2$, the NH (Normal Hierarchy) and IH (Inverted Hierarchy) refer to the two possible mass heirarchy of the light neutrino masses. Values not explicitly labeled as NH or IH are common to both hierarchies.

The light-neutrino masses for NH (IH) can be parameterized in term of the lightest neutrino mass $m_{lighttest} = m_1 (m_3)$. 
For NH $m_1< m_2 <m_3$ such that      
\bea
m_2=\sqrt{m_{lightest}^2 +\Delta m_{21}^2 } ,\quad
m_3=\sqrt{m_{lightest}^2 +\Delta m_{21}^2+\Delta m_{32}^2} ,
\label{eq:mlightNH}
\eea
while for the IH $m_3< m_1 <m_2$ with 
\bea
m_1=\sqrt{m_{lightest}^2 - \Delta m_{32}^2-\Delta m_{21}^2 } ,\quad
m_2=\sqrt{m_{lightest}^2 -\Delta m_{32}^2 }. 
\label{eq:mlightIH}
\eea
Using Eqs.~(\ref{eq:seesaw}) and~(\ref{eq:Dnu}), the Dirac mass matrix can be expressed as 
\bea
m_D = \sqrt{M_N} O \sqrt{D_\nu} U_{\rm MNS}^\dag,  
\label{eqn:CIR}
\eea
where $O$ is an arbitrary complex orthogonal matrix satisfying $O^T O = \mathbf{1}$ \cite{Casas:2001sr,Ibarra:2003up}. The explicit forms of the matrices $O$, $\sqrt{M_N}$, and $\sqrt{D_\nu}$ depend on the number of RHN generations. 
We will now separately discuss the cases with two generations (minimal seesaw) and three generations of RHNs.

\subsection{SM + 2 RHNs}
This is the minimal seesaw needed to reproduce the neutrino oscillation observations. 
Both $\sqrt{M_N}$ and $O$ are $2 \times 2$ matrices,  
\bea
\sqrt{M_N} &=& \left(\begin{matrix}
\sqrt{m_{N^1}} & 0\\
0 & \sqrt{m_{N^2}}  
\end{matrix}\right), 
\nonumber \\
O
&=&
\left(\begin{matrix}
\cosh (a + i b) & \sinh (a + i b) \\
-\sinh (a + i b)  & \cosh (a+ i b)
\end{matrix}\right), 
\eea
where
$a$ and $b$ are real valued parameters. 
For NH and IH, the matrix $\sqrt{D_\nu}$ is given by 
\bea
\sqrt{D_\nu^{\rm NH}} &=& \left(\begin{matrix}
0 & \sqrt{m_2} & 0\\
0 & 0 & \sqrt{m_3}  
\end{matrix}\right),\\
\sqrt{D_\nu^{\rm IH}} &=& \left(\begin{matrix}
\sqrt{m_1} & 0 & 0\\
0 & \sqrt{m_2} & 0  
\end{matrix}\right),    
\eea
respectively. This $3\times 3$ diagonal mass eigenvalue matrix is given by $ \sqrt{D_\nu}^T\sqrt{D_\nu} = D_\nu  = {\rm diag} (m_1, m_2, m_3)$.  
In the minimal seesaw, $m_{lightest} = 0$. 
Using Eq.~(\ref{eq:mlightNH}) for NH with $m_{lightest} = m_1 = 0$, we obtain $m_2 = 8.71\times 10^{-3}$ eV and $m_3 = 5.00\times 10^{-2}$ eV.  
Using Eq.~(\ref{eq:mlightIH}) for IH with $m_{lightest} = m_3 = 0$, we obtain $m_1 (m_3) = 4.85 \times 10^{-2}$ eV and $m_2 = 4.92\times 10^{-2}$ eV. 

\subsection{SM + 3 RHNs }
For the case with three RHNs, the neutrino mass matrices are given by   
\bea
\sqrt{M_N} &\equiv& \mbox{diag}(\sqrt{m_{N^1}},\sqrt{m_{N^2}},\sqrt{m_{N^3}}), 
\nonumber \\
\sqrt{D_\nu^{\rm NH}} &=& {\rm diag} \left(\sqrt{m_{lightest}}, \sqrt{m_2}, \sqrt{m_3}\right),\\
\sqrt{D_\nu^{\rm IH}} &=& {\rm diag} \left(\sqrt{m_1}, \sqrt{m_2}, \sqrt{m_{lightest}}\right). 
\eea
The complex $3 \times 3$ orthogonal matrix can be expressed as $O = O_1 \; O_2 \;O_3$, where $O_i$ is given by 
\bea
O_1
&=&
\left(\begin{matrix}
1 & 0 & 0\\
0 & \cos ( a_{23} + i b_{23} ) & \sin ( a_{23} + i b_{23} )\\
0 & -\sin ( a_{23} + i b_{23} ) & \cos ( a_{23} + i b_{23} )
\end{matrix}\right), \nonumber \\
O_2
&=&
\left(\begin{matrix}
\cos(a_{13} + i b_{13}) & 0 & \sin(a_{13} + i b_{13})\\
0 & 1 & 0\\
-\sin(a_{13} + i b_{13}) & 1 & \cos(a_{13} + i b_{13})
\end{matrix}\right) , \nonumber \\
O_3
&=&
\left(\begin{matrix}
\cos(a_{12} + i b_{12}) & \sin(a_{12} + i b_{12}) & 0\\
- \sin(a_{12} + i b_{12}) & \cos(a_{12} + i b_{12}) & 0\\
0 & 0 & 1
\end{matrix}\right), 
\eea
where $a_{ij}$ and $b_{ij}$ are real valued parameters.

\section{Analytic Formula for $\epsilon$}
CP asymmetric decay of RHN originates from the interference between the tree and one-loop decay amplitudes. 
This is parametrized by the $CP$ asymmetry parameter $\epsilon_i$ defined as \cite{Covi:1996wh, Buchmuller:1997yu}  
\bea
\epsilon_i&\equiv&\frac{
\sum_j\left[\Gamma (N_R^i\rightarrow\ell_L^j+H)
-\Gamma(N_R^i\rightarrow\overline {\ell_L^j}+H^\dag)\right]}{\sum_j\left[\Gamma (N_R^i\rightarrow\ell_L^j+H)
+\Gamma(N_R^i\rightarrow\overline {\ell_L^j}+H^\dag)\right]},\nonumber \\
&=&
- \frac{1}{4\pi v_h^2} \frac{m_{N^i}}{(m_Dm_D^\dag)_{ii}} \sum_{j\neq i}\frac{ {\rm Im}[(m_D m_D^\dag)^2_{ij}]}{m_{N^j}}\left( \frac{1}{2}V_j+S_j\right),
\label{eqn:epsilon}
\eea
where $ \Sigma_j \Gamma(N_R^i\rightarrow\ell_L^j+H)\equiv \Gamma_i $ is the total decay width of $N_R^i$ to the SM final states and $V_j$ and $S_j$ are the vertex and self-energy corrections, respectively: 
\bea
\Gamma_i &=& \frac{m_{N^j}}{8\pi} (Y_D Y_D^\dagger)_{ii}  =  \frac{m_{N^j}}{4\pi v_h^2} (m_D m_D^\dag)_{ii}\; , \nonumber\\
V_j&=& 2\frac{m_{N^j}^2}{m_{N^i}^2}\left[\left(1+\frac{m_{N^j}^2}{m_{N^i}^2}\right)\ln\left(1+\frac{m_{N^i}^2}{m_{N^j}^2}\right)-1\right],
\nonumber\\
S_j&=&\frac{m_{N^j}^2\Delta M_{ji}^2}{(\Delta M_{ji}^2)^2+m_{N^i}^2\Gamma_j^2}\; ,
\label{eq:epsilon2}
\eea
with $\Delta M_{ji}^2=m_{N^j}^2-m_{N^i}^2$. 
If the masses are degenerate, $m_{N^i} \simeq m_{N^j}$ such that $\Delta M_{ji}^2 \gtrsim m_{N^i} \Gamma_j$, then $S_i \gg 1$ leading to a significant enhancement of  $\epsilon$. This scenario is known as resonant leptogenesis \cite{Flanz:1996fb, Pilaftsis:1997jf}. 
In the following, we will consider RHN masses to be hierarchical, in which case both corrections $V_j, S_j \simeq 1$.

Using the expression for the Dirac mass matrix $m_D$ in Eq.~(\ref{eqn:CIR}), we obtain 
\bea
(m_D m_D^\dag)_{ij} =  \sqrt{m_{N^i} \; m_{N^j}} \left( O \tilde{D_{\nu}} O^\dagger \right)_{ij}, 
\eea
where $\tilde{D_{\nu}} = \sqrt{D_\nu} {\sqrt{D_\nu}}^\dagger$ and we have exploited the unitarity of the MNS matrix $U^\dagger U = 1$. 
This simplifies the expression for the decay width and the $CP$ asymmetry parameter, 
\bea
\Gamma_i &=& \; = \frac{3 m_{N^i}^2 }{8\pi v_h^2} \left( O {\tilde D_\nu} O^\dagger \right)_{ii}, \nonumber \\
\epsilon_i \; &=&
- \frac{3}{8\pi v_h^2} \frac{ m_{N^i}}{{(O {\tilde D_\nu} O^\dagger)_{ii}}}\sum_{j\neq i}\;  {{\rm Im}[(O {\tilde D_\nu} O^\dagger)^2_{ij}]}. 
\label{eqn:epsilon1}
\eea
Note that both the CP asymmetry parameter $\epsilon_i$ and the decay width $\Gamma_i$ associated with the decay of $N_i$ are independent of the MNS matrix but depends on the neutrino mass eigenvalues.

In this work, we focus on a hierarchical mass spectrum for the RHNs with $N_R^1$ being the lightest. 
In thermal leptogenesis, even all $N_{R^i}$ are in thermal equillibrium in the early universe, the out-of-equilibrium decay of $N_{R^1}$ determines the resultant lepton asymmetry since the lepton asymmetry  generated by heavier RHNs are washed out by $N_{R^1}$ which is still in thermal equilibrium. 
In our discussion about non-thermal leptogenesis, we simply assume that only $N_{R^1}$ is created by a decay of a particle (inflaton), either due to kinematic constraints or the nature of the couplings involved.  
We now present analytic expressions for the decay width and the CP asymmetry parameter associated with $N_R^1$, using the explicit forms of the matrices $O$, ${\tilde D_\nu}$, and $\sqrt{M_N}$ introduced in the previous section.

\subsection{SM + 2 RHNs}
For the NH with $m_{lightest} = m_1 = 0$, we obtain 
\bea
\Gamma_1^{NH} &=& 
\left(\frac{m_{N^1}^2}{8\pi v_h^2} \right) \left((m_2-m_3)\cos2a+(m_2+m_3) \cosh2 b\right), 
%%%%%%%%%%%%%%%%%%%
\nonumber \\
%%%%%%%%%%%%%%%%%%
\epsilon_1^{NH} &=& \left(\frac{3 m_{N^1}}{8\pi v_h^2}\right)\frac{\left(m_2^2 - m_3^2\right) \sin2 a \times \sinh2b}{\left(m_2 -m_3\right) \cos2a  +  \left(m_2 +m_3\right)\cosh2b}. 
\label{eq:SM2epsNH}
\eea
For the IH with $m_{lightest} = m_3 = 0$, we obtain 
\bea
\Gamma_1^{IH} &=& \left(\frac{m_{N^1}^2}{8\pi v_h^2} \right) \left((m_1-m_2)\cos2a+(m_1+m_2) \cosh2b\right), 
%%%%%%%%%%%%%%%%%%%
\nonumber \\
%%%%%%%%%%%%%%%%%%
\epsilon_1^{IH} &=& 
\left(\frac{3 m_{N^1}}{8\pi v_h^2}\right) \frac{\left(m_1^2 - m_2^2\right) \sin2 a \times \sinh2b}{\left(m_1 -m_2\right) \cos2a  +  \left(m_1 +m_2\right)\cosh2b}. 
\label{eq:SM2epsIH}
\eea

\subsection{SM + 3 RHNs }
For $N_R^1$ decay, 
\bea
\Gamma_1 &=& \; \frac{3 m_{N^1}^2 }{8\pi v_h^2} \left( O D_\nu O^\dagger \right)_{11}, \nonumber \\
\epsilon_1 &=&
- \frac{3 m_{N^1}}{8\pi v_h^2}\; 
\frac{\sum_{j= 2,3}{\rm Im}[(O D_\nu O^\dagger)^2_{1j}]}{(O D_\nu O^\dagger)_{11}},
\label{eq:SM3eps}
\eea
where   
\bea
4\times \sum_{j=2,3}{\rm Im}[(O D_\nu O^\dagger)^2_{1j}] \; =&& 2\; \Delta m_{32}^2 \sin 2 a_{13} \sinh 2 b_{13} \nonumber \\ 
&+& \Delta m_{21}^2 \Big(1+ \cos 2 a_{12} \cosh 2 b_{12} \Big) \sin 2 a_{13} \sinh 2 b_{13}\nonumber \\
&+& \Delta m_{21}^2 \Big(1+ \cos 2 a_{13} \cosh 2 b_{13}\Big) \sin 2 a_{12} \sinh 2 b_{12},
%%%%%%%%%%%%%%%%%%%
\nonumber \\ 
%%%%%%%%%%%%%%%%%%
4\times (O D_\nu O^\dagger)_{11} \; = &&
2 \;m_3 \Big( \cosh 2b_{13} - \cos 2a_{13}\Big) 
\nonumber \\
&+& m_1 \Big(\cosh 2b_{12} + \cos 2 a_{12}\Big) \Big(\cosh 2b_{13}+ \cos 2a_{13}\Big) 
\nonumber \\
&+& m_2 \Big(\cosh 2b_{12} -\cos 2 a_{12}\Big) \Big(\cosh 2b_{13} + \cos 2a_{13}\Big). 
\eea
%%%%%%%%%%%%%%%%%%%%%%%%%%%%%%%%%%%%%%%%%
where $\Delta m_{ij}^2 \equiv m_i^2 - m_j^2$. Note that $\epsilon_1$ is independent of $a_{23}$ and $b_{23}$. Also, we can see that $\epsilon_1 = 0$ if either $(a_{12} = a_{13} = 0)$ or $(a_{13} = b_{12} = 0)$ or $(a_{12} = b_{13} = 0)$.

%For $a_{12}= 0, b_{12} = 0, a_{13} = \frac{\pi}{4}$, 
%\bea
%\epsilon_1 &=& \left(\frac{3 m_{N^1}}{8\pi v_h^2}\right)\left(m_2 -m_3\right)\tanh2b_{13} 
%\eea
%which is the same as the two generation with NH. 
%For $a_{12}= \frac{\pi}{4}, b_{12} = 0, a_{13} = 0$, 
%\bea
%\epsilon_1 &=& \left(\frac{3 m_{N^1}}{8\pi v_h^2}\right)\left(m_1 -m_2\right)\tanh2b_{12} 
%\eea
%which is the same as the two generation with NH. 

%%%%%%%%%%%%%%%%%%%%%%%%%%%%%%%%%%%%%%%
\section{Leptogenesis}
%%%%%%%%%%%%%%%%%%%%%%%%%%%%%%%%%%%%%%%
Depending on the thermalization history of the RHNs, leptogenesis can be broadly categorized into two types: thermal and non-thermal. In thermal leptogenesis, the RHNs are in thermal equilibrium with the SM plasma in the early universe. 
In contrast, in non-thermal leptogenesis, the RHNs are never thermalized. 
We now discuss each scenario in detail.

%%%%%%%%%%%%%%%%%%%%%%%%%%%%%%%%%%%%%%%
\subsection{Thermal Leptogenesis}
%%%%%%%%%%%%%%%%%%%%%%%%%%%%%%%%%%%%%%%
In thermal leptogenesis with hierarchical RHN masses, the final lepton asymmetry is determined by the lepton asymmetry generated by the lightest RHN, $N_R^1$, decay. This can be evaluated by solving the coupled Boltzmann equations for number densities of leptons ($L$) and $N_R^1$ \cite{Plumacher:1996kc} 
\bea
{d Y_{1}\over dz} & = & - \frac{z}{s(m_{N^1}) H(m_{N^1})}\left(\frac{Y_{1}}{Y_{1}^{eq}} -1\right) \; (\gamma_D+\gamma_S)\, \;, \label{lg1} \nonumber \\
{dY_{L}\over dz} & = & \frac{z}{s(m_{N^1}) H(m_{N^1})} \left[\left(\frac{Y_{1}}{Y_{1}^{eq}} -1\right) \; \epsilon_1\,\gamma_D - \, \left(\frac{Y_{L}}{Y_{\ell}^{eq}}\right) \,\gamma_{W} \right]\;. 
\label{eq:boltzmann}
\eea
Here, $z = \frac{m_{N^1}}{T}$ is a dimensionless parameter, $Y= n/s$ is the yield (ratio of number density and entropy density), the label ``$eq$" identifies equilibrium values, and $H(m_{N^1})$ and $s(m_{N^1})$ are the Hubble parameter and entropy density parameters, respectively, evaluated at $T = m_{N^1}$ 
\bea
H (m_{N^1}) &=& \left(\frac{\pi^2 g_*^{SM}}{90}\right)^{1/2} \frac{m_{N^1}^2}{M_P}, \nonumber \\
s(m_{N^1}) &=& \frac{2\pi^2}{45} g_*^{SM} m_{N^1}^3, 
\label{eq:Hs}
\eea
and $g^{SM}_*$ is the relativistic degree of freedom evaluated at $T = m_{N^1}$ which we fix to be $g^{{SM}}_* = 106.75$. 
The $\gamma$'s denote thermally averaged rates for interactions involving $N_R^1$ and SM particles:  
$\gamma_D$ is the decay rate of $N_R^1$,  
$\gamma_S$ includes SM Higgs boson ($h$) mediated $t$ and $s$ channel scattering of $N_R^1$,  
\begin{equation}
\gamma_S=2\gamma_{h,t}^{(N_R^1)}+4\gamma_{h,s}^{(N_R^1)}\;, 
\end{equation}
and $\gamma_W$ includes the inverse processes which can wash out of yield of lepton number,  
\begin{equation}
\gamma_W={1\over 2}\gamma_{D}
+2\left(\gamma_N^{(l)}+ \gamma_{N,t}^{(l)} +\gamma_{h,t}^{(l)}\right) + 
\left({Y_{1}\over Y_{1}^{\rm eq}}\right) \gamma_{h,s}^{(l)} , 
\end{equation}
where $\gamma_{h,t}$ and $\gamma_{h,s}$ are scattering processes which violates lepton number by 1 unit ($\Delta L=1$) while $\gamma_{h,t}$ and $\gamma_{h,s}$ are  scattering processes which violates lepton number by 2 units ($\Delta L=2$).

The Boltzmann equation can be solved numerically to evaluate the yield of lepton number. Taking into account the sphaleron process, the resulting baryon asymmetry for an initial condition $Y_{L}(0) = 0$ is given by 
\bea
Y_B = -C_{sph} \times Y_L (z= \infty). 
\eea
\begin{figure}[t!]
    \centering
\includegraphics[width=0.5\linewidth]{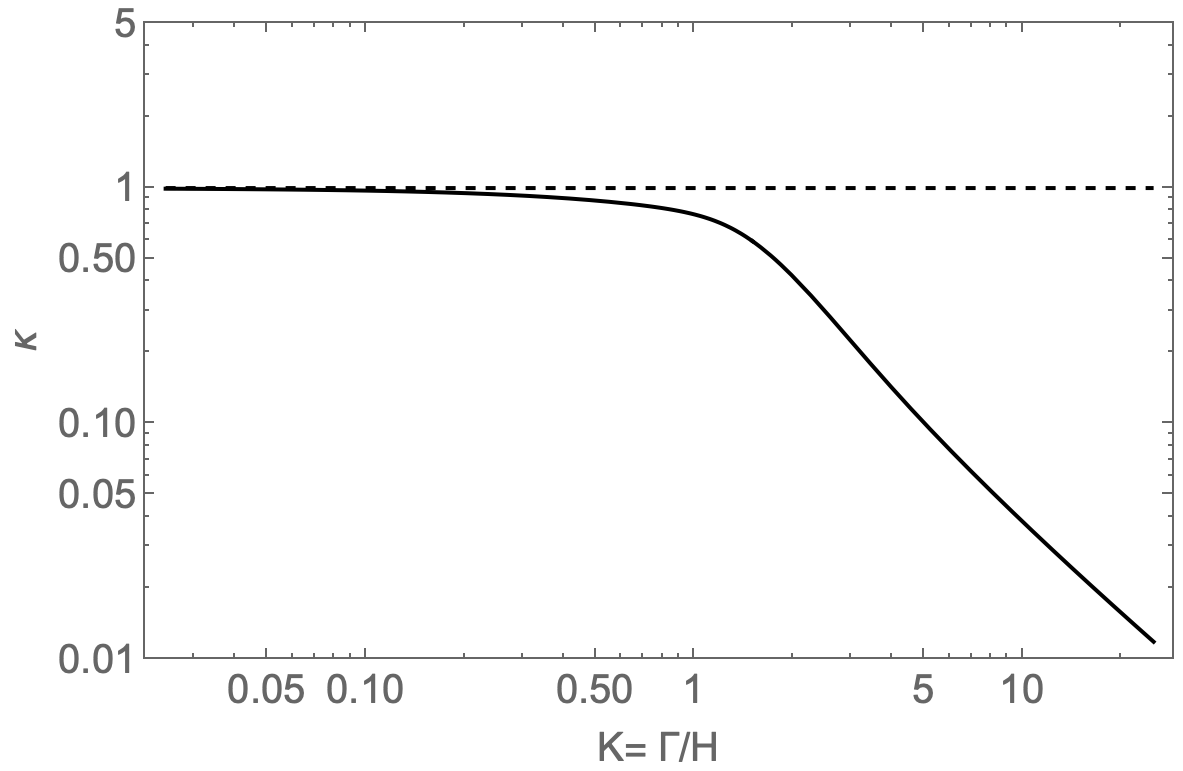}
    \caption{Plot of efficiency factor $\kappa$ as a function of $K= \Gamma/H$}
    \label{fig:EF}
\end{figure}

The Boltzmann equation can also be solved analytically. By neglecting the contribution from scattering processes, which are known to be subdominant compared to decay and inverse decay processes, the resulting baryon asymmetry is given by   
\bea
\frac{n_B}{s} = -C_{sph} \times \frac{n_{L}}{s}= -C_{sph}  \frac{\epsilon_1 \times n_{N^1} (m_{N^1})}{s (m_{N^1})} \times \kappa = -C_{sph}\times \frac{135 \;\zeta (3)}{4\pi ^4} \; \frac{1}{g^{SM}_*}\; \epsilon_1  \times \kappa  \;, 
\label{eq:epsthermal}
\eea
Here, $n_L/s$ is evaluated using $n_L = \epsilon_1 n_{N^1}^{\text{eq}}$ and $s$ at $T = m_{N^1}$, with the efficiency factor $\kappa$ included to account for the dilution of the lepton asymmetry due to inverse decay processes. 
The analytic expression for $\kappa$ is obtained in Ref.~\cite{Buchmuller:2004nz}, 
\bea
\kappa (K) &\simeq&
{2\over z_{\rm B}(K)K}\left(1-e^{-{1\over 2}z_{\rm B}(K)K}\right), 
\nonumber \\ 
z_B(K) &\simeq& 1+{1\over2}\,\mbox{ln}\left(1+{\pi K^2\over1024} 
\left[\mbox{ln}\left({3125\pi K^2\over1024}\right)\right]^5\right), 
\nonumber \\
K &=& \frac{\Gamma_1}{H (m_{N^1})}. 
\label{eq:dilution}
\eea
Figure~\ref{fig:EF} shows the efficiency factor as a function of $K$, demonstrating that baryon asymmetry is substantially diluted in the strong washout regime ($K \gg 1$). $\kappa$ approaches 1 in the limit $K \to 0$, which means the absence of washout.

Using Eq.~(\ref{eqn:epsilon1}), we can express $\kappa$ as 
\bea
K = \frac{\Gamma_1}{H (m_{N^1})} \; = \;  \frac{1}{4 \pi^2} \; \sqrt{\frac{90}{g_*^{SM}}} \; \frac{M_P}{v_h^2}\; (O \tilde {D_\nu} O^\dagger)_{11}.  
\label{eq:dilution1}
\eea
Note that both $K$ and hence $\kappa$ are independent of $m_{N^1}$, whereas the CP asymmetry parameter scales as $\epsilon_1 \propto m_{N^1}$. Hence, the baryon-to-entropy ratio $n_B/s \propto m_{N^1}$.  
Requiring $n_B/s \simeq 8.7 \times 10^{-11}$ to match the observed baryon asymmetry, we now evaluate the minimum value of the lightest RHN mass, $m_{N^1}$, for the scenarios with two and three generations of RHNs.

%%%%%%%%%%%%%%%%%%%%%%%%%%%%%%%%%%%%%%%
\subsubsection{Results: SM + 2 RHNs}
%%%%%%%%%%%%%%%%%%%%%%%%%%%%%%%%%%%%%%%
Since $m_{\text{lightest}} = 0$ in the minimal seesaw scenario, all light neutrino masses are fixed. 
Consequently, $n_B/s$ only depends on the $m_{N^1}$ and the two real parameters $a$ and $b$. 
The explicit formulas for $\epsilon_1$ and $\Gamma_1$, expressed in terms of $a$ and $b$, are given in Eqs.~(\ref{eq:SM2epsNH}) and (\ref{eq:SM2epsIH}) for NH and IH, respectively.

\begin{figure}[t!]
    \centering
\includegraphics[width=0.31\linewidth]{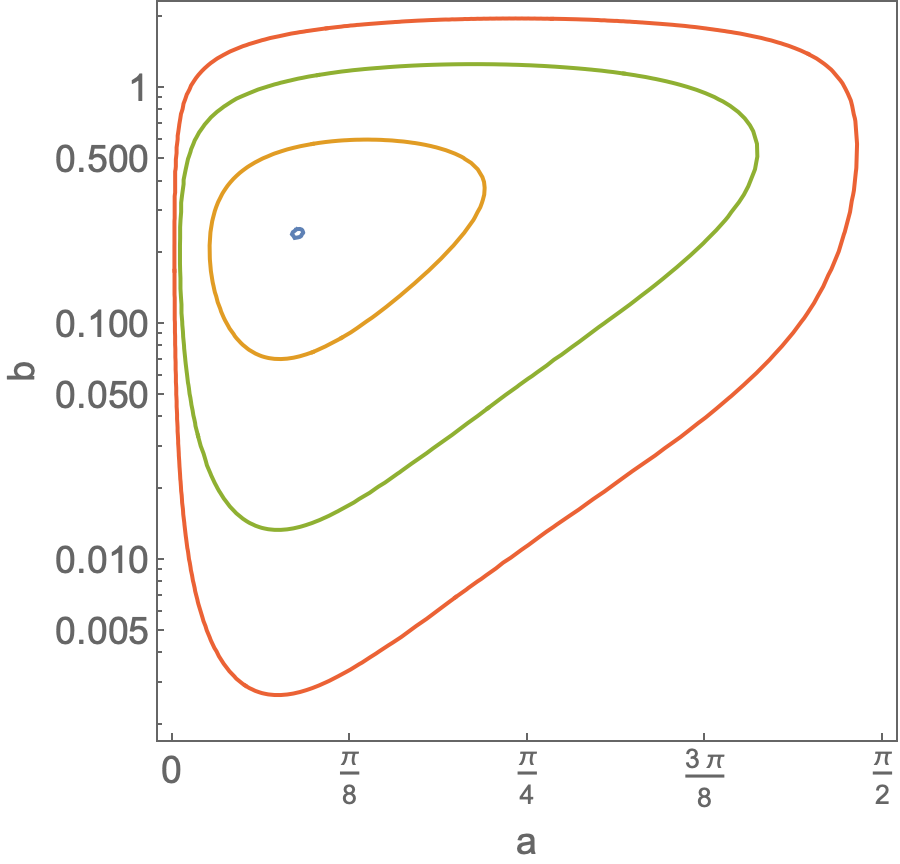}\;\;\; 
\includegraphics[width=0.31\linewidth]{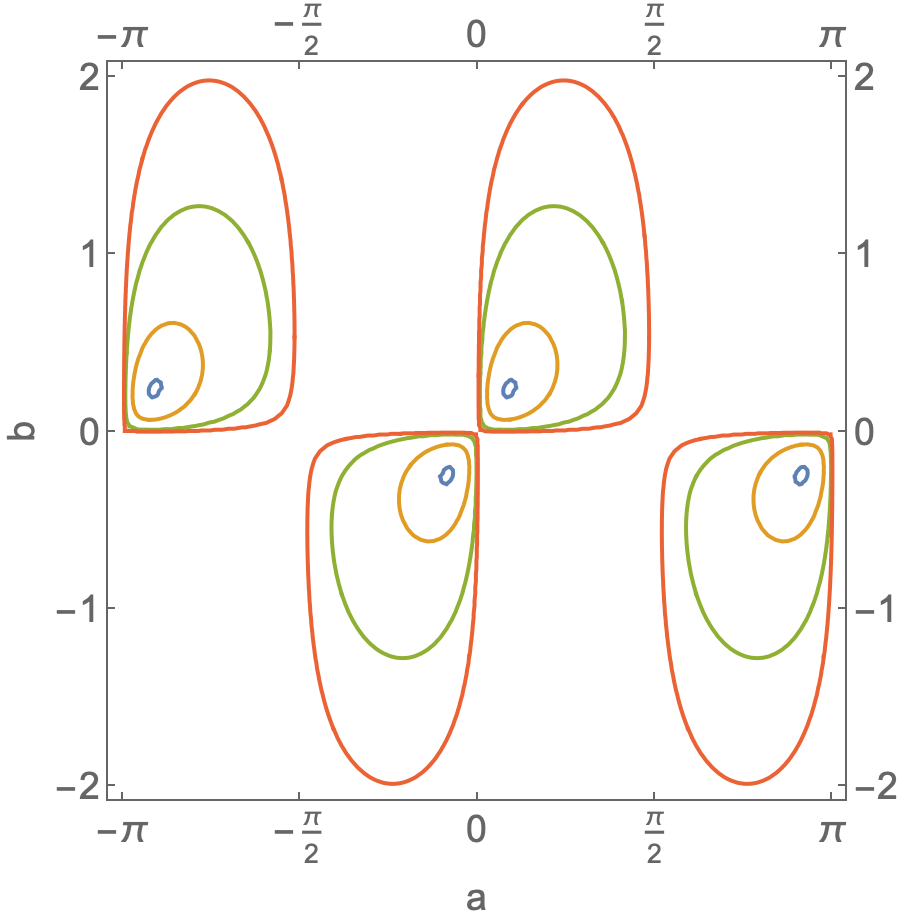}\;\;\; 
\includegraphics[width=0.31\linewidth]{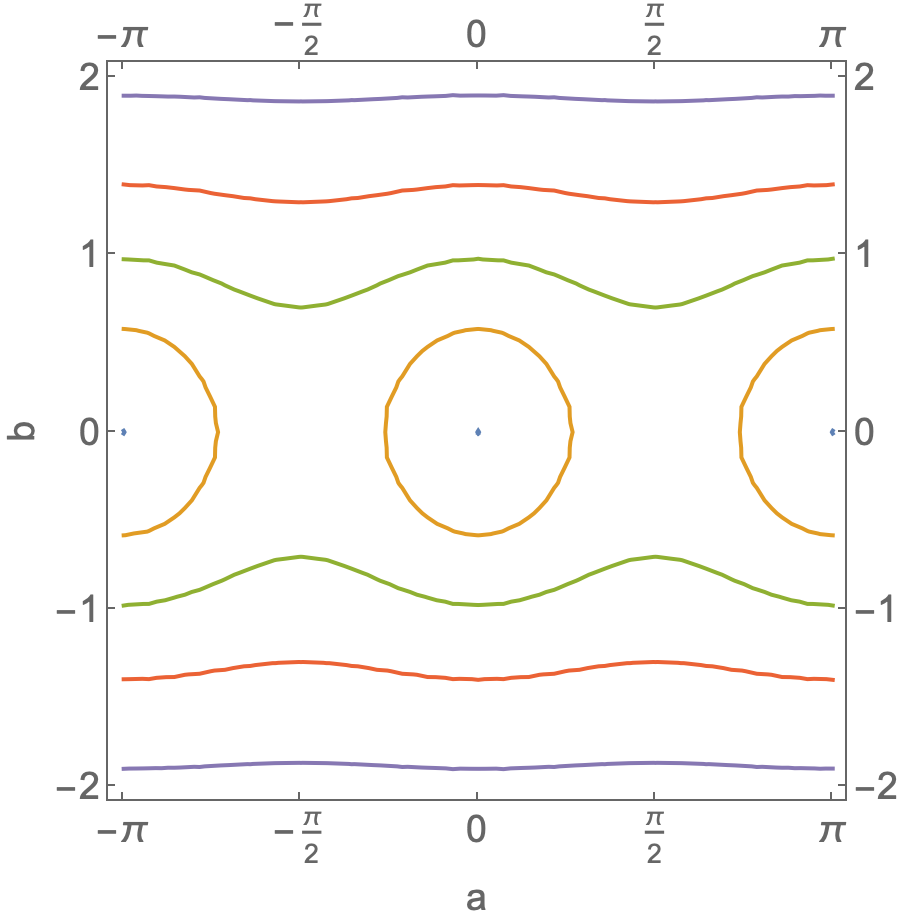}
    \caption{Results for the NH. The first two panels show contours of fixed $m_{N^1}$ that yield the observed baryon asymmetry, with the innermost corresponding to $m_{N^1}^{{min}} \simeq 6.24 \times 10^{10}$ GeV, and the outer contours representing $2$, $10$, and $50$ times this value. The right panel shows contours of fixed $\kappa = 5.1 \times 10^{-2}, 10^{-2}, 3 \times 10^{-3}, 10^{-3}$, and $3 \times 10^{-4}$ (innermost to outermost). }
    \label{fig:SM2thNH}
\end{figure}

In Fig.~\ref{fig:SM2thNH}, we show the result for the NH. The left panel shows contours for fixed $m_{N^1}$ in the $a$–$b$ plane along which the observed baryon asymmetry is produced. The innermost contour marks the minimum value $m_{N^1}^{min} \simeq 6.24 \times 10^{10}$ GeV, with outer contours at $2$, $10$, and $50$ times this minimum. 
The middle panel shows that the result for a wider range of $a$ and $b$ with contours corresponding to $1.03$, $2$, $10$, and $50$ times the $m_{N^1}^{min}$. 
The right panel shows contours for fixed $\kappa = 5.1 \times 10^{-2}, 10^{-2}, 3 \times 10^{-3}, 10^{-3}$, and $3 \times 10^{-4}$ (from innermost to outermost). 
Since $\kappa$ is independent of $m_{N^1}$, the plot illustrates that larger values of $b$ lead to stronger washout, necessitating a higher $m_{N^1}$ to reproduce the observed asymmetry.

\begin{figure}[th!]
    \centering
\includegraphics[width=0.31\linewidth]{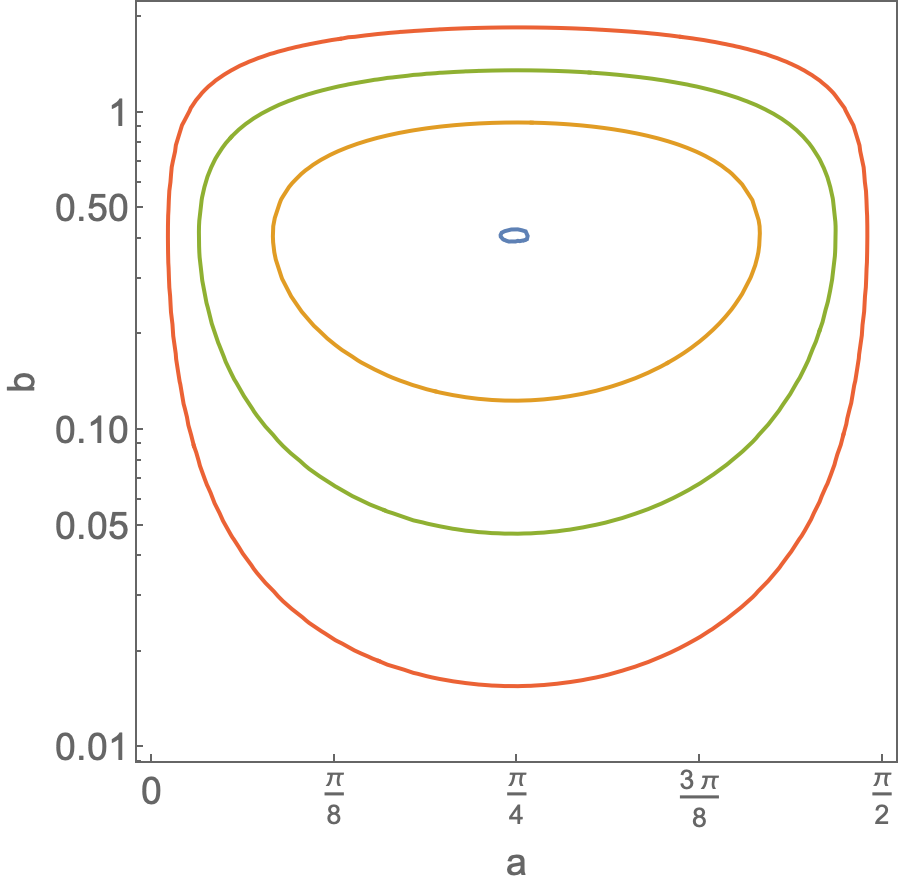}\;\;\; 
\includegraphics[width=0.31\linewidth]{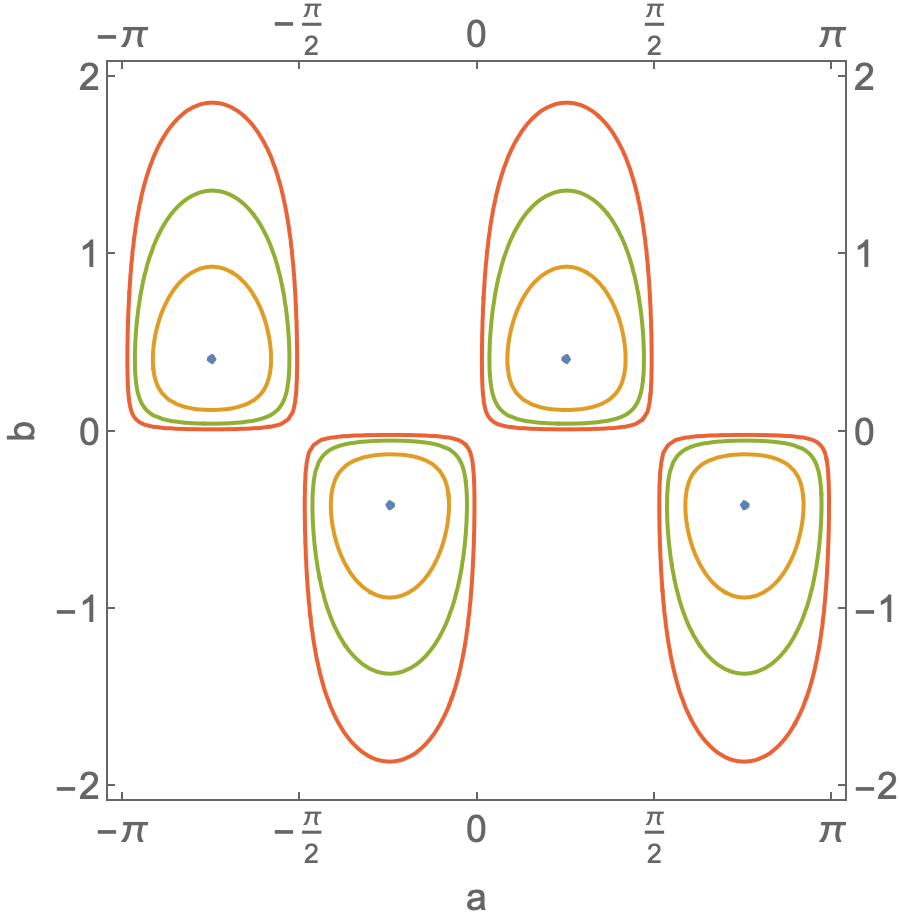}\;\;\; 
\includegraphics[width=0.31\linewidth]{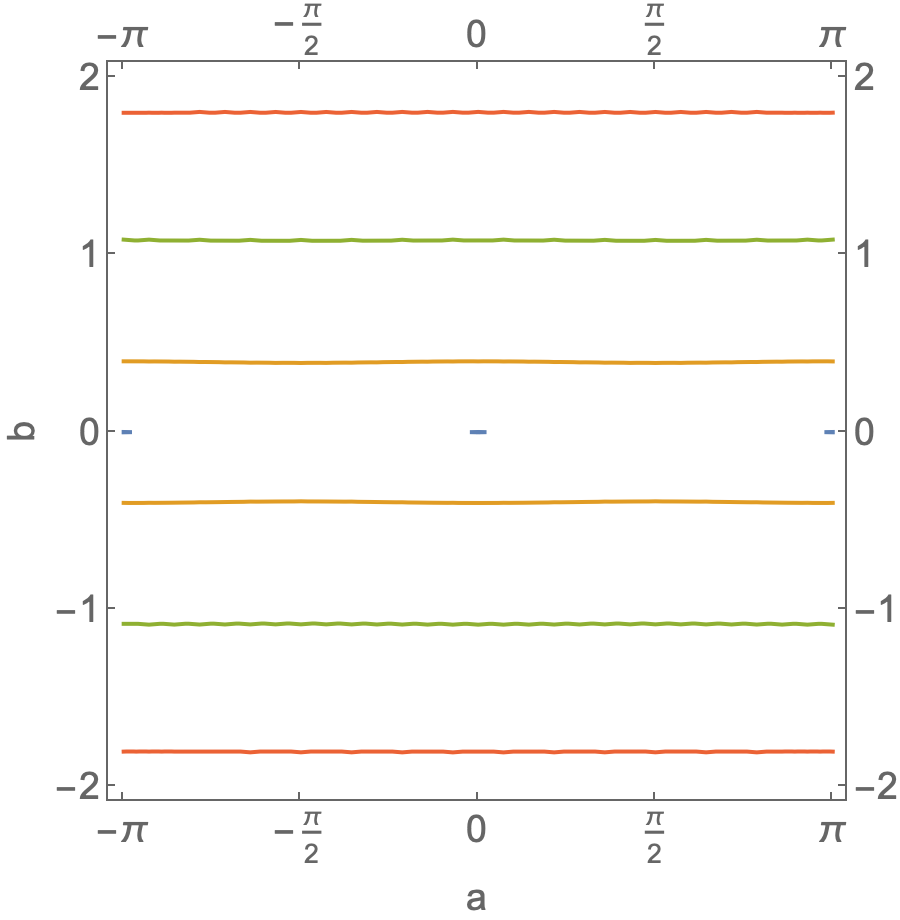}
    \caption{Results for the IH. The first two panels show contours of fixed $m_{N^1}$ that yield the observed baryon asymmetry, with the innermost corresponding to $m_{N^1}^{{min}} \simeq 1.62 \times 10^{13}$ GeV, and the outer contours representing $2$, $5$, and $15$ times this value. The right panel shows contours for fixed $\kappa = 5.7 \times 10^{-3}, 4.0\times 10^{-3},  10^{-3}$, and $2 \times 10^{-4}$ (innermost to outermost). }
    \label{fig:SM2thIH}
\end{figure}

Figure~\ref{fig:SM2thIH}, shows the result for the IH. 
The left panel shows contours for fixed $m_{N^1}$ with the innermost contour marking the minimum value $m_{N^1}^{min} \simeq 1.62 \times 10^{13}$ GeV, and the outer contours corresponding to $2$, $5$, and $15$ times the $m_{N^1}^{min}$.
The middle panel shows contours for the same $m_{N^1}$ values as in the left panel, but over a wider range of $a$ and $b$ values. 
The right panel shows contours for fixed $\kappa = 5.7 \times 10^{-3}, 4.0\times 10^{-3},  10^{-3}$, and $2 \times 10^{-4}$ (from innermost to outermost). 
Comparing with the corresponding plots for the NH, we find that the dilution is more efficient for IH which leads to larger value of $m_{N^1}^{min}$. 

%%%%%%%%%%%%%%%%%%%%%%%%%%%%%%%%%%%%%%%
\subsubsection{Results: SM + 3 RHNs}
%%%%%%%%%%%%%%%%%%%%%%%%%%%%%%%%%%%%%%%
For three generations of RHNs, the light neutrino masses are determined up to the value of the lightest neutrino mass, $m_{{lightest}}$, according to Eqs.~(\ref{eq:mlightNH}) and  (\ref{eq:mlightIH}) for NH and IH, respectively.
Together with the explicit expressions for $\Gamma_1$ and $\epsilon_1$ in Eq.~(\ref{eq:SM3eps}), the baryon asymmetry is determined by $m_{\text{lightest}}$ and the real parameters $a_{12}$, $a_{13}$, $b_{12}$, and $b_{13}$.
The analytical formulas allows us to perform a parameter scans over all the parameters in a very short time to evaluate the $m_{N^1}^{min}$.

\begin{figure}[t]
    \centering
\includegraphics[width=0.45\linewidth]{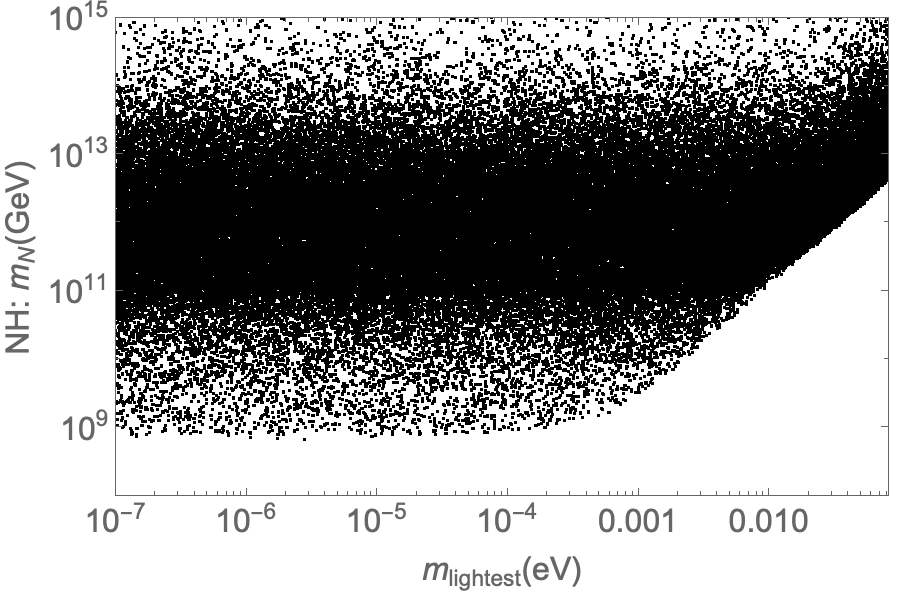}\qquad \;
\includegraphics[width=0.47\linewidth]{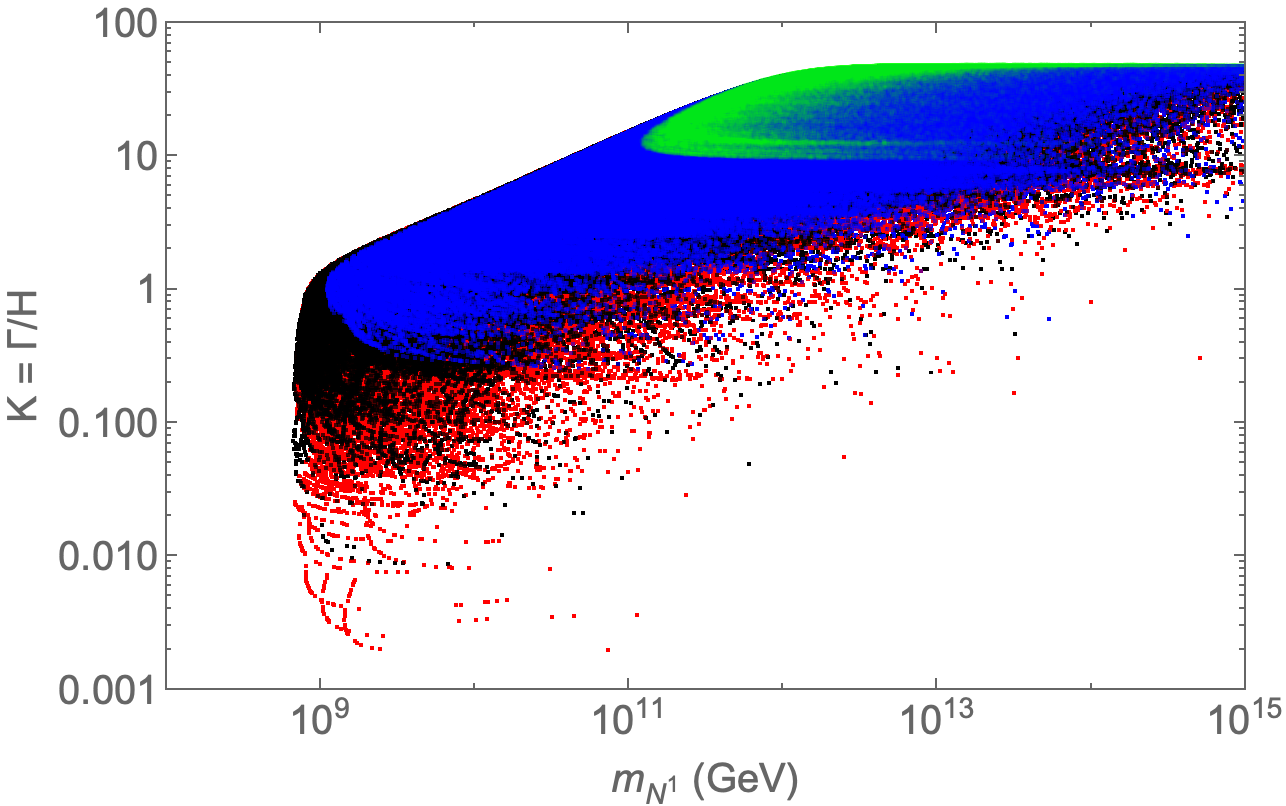}
    \caption{The two plots shows results from a parameter scan for the NH. The left panel displays the minimum values of $m_{N^1}$ for various values of $m_{{lightest}}$ obtained by scanning over all other parameters. We find that the smallest value of $m_{N^1}^{min} = 6.68\times 10^8$ GeV. The right panel shows $K = \Gamma_1/H$ for different $m_{N^1}$ values that reproduce the observed baryon asymmetry. The green, blue, black, and red points correspond to $m_{{lightest}}/\text{eV} = 10^{-2}, 10^{-3}, 10^{-4}$, and $10^{-8}$, respectively.}
    \label{fig:SM3thNH}
\end{figure}

\begin{figure}[t!]
    \centering
\includegraphics[width=0.45\linewidth]{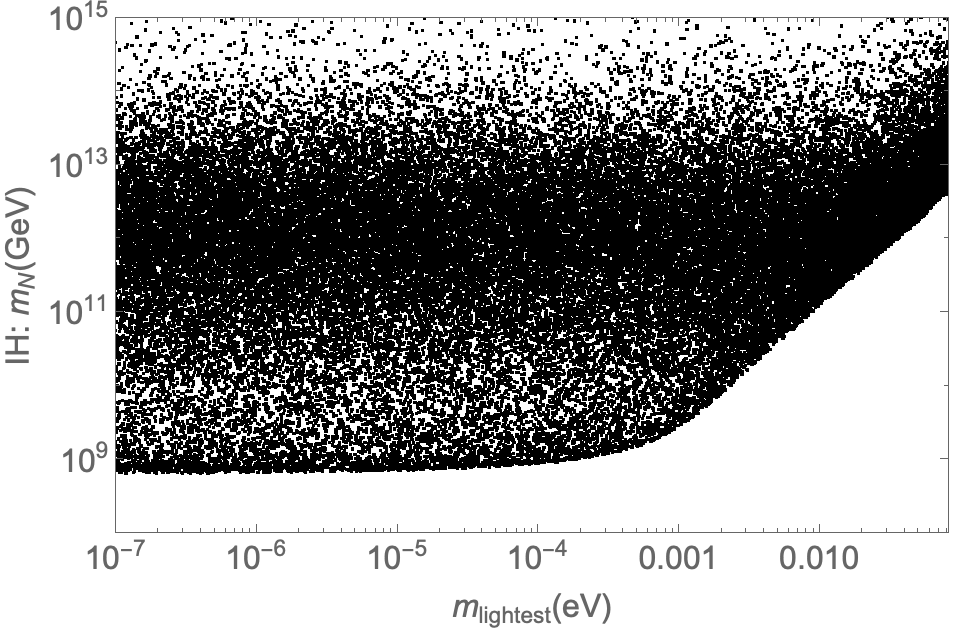}\qquad \;
\includegraphics[width=0.47\linewidth]{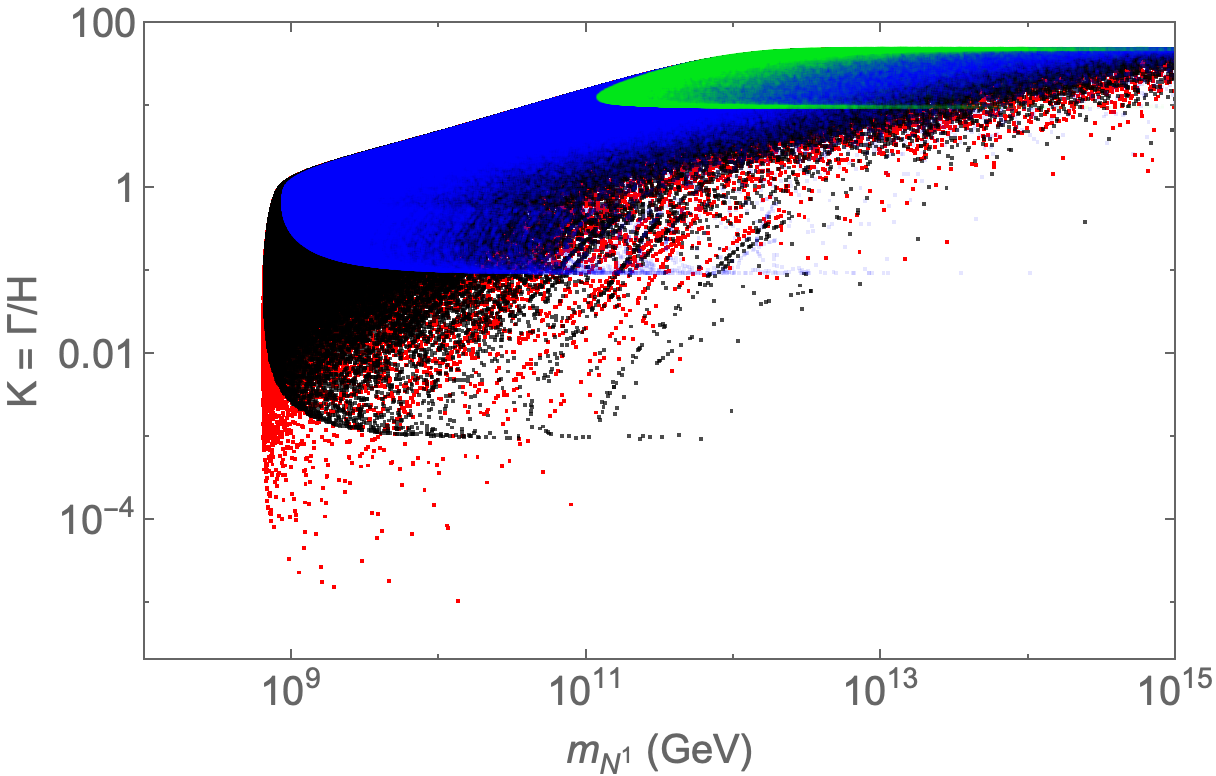}
    \caption{The two plots shows results from a parameter scan for the IH. The left panel displays the minimum values of $m_{N^1}$ for various values $m_{{lightest}}$ obtained by scanning over all other parameters. We find that the smallest value of $m_{N^1}^{min} = 6.52\times 10^8$ GeV. The right panel shows $K = \Gamma_1/H$ for different $m_{N^1}$ values that reproduce the observed baryon asymmetry. The green, blue, black, and red points correspond to $m_{{lightest}}/\text{ eV} = 10^{-2}, 10^{-3}, 10^{-4}$, and $10^{-8}$, respectively.}
    \label{fig:SM3thIH}
\end{figure}

We show our result for NH in Fig.~\ref{fig:SM3thNH}.  
The left panel shows the minimum values of $m_{N^1}$ obtained for various $m_{lightest}$ after scanning over remaining parameters.  
We find that the minimum value of $m_{N^1}$ is independent of $m_{lightest}$ for $m_{lightest} \lesssim 10^{-3}$ eV. 
We obtain the smallest value of $m_{N^1}^{min} = 6.68\times 10^8$ GeV.
In the right panel, we show results for the scan performed for fixed values of $m_{{lightest}}$. It shows $K = \Gamma_1/H$ for various values of $m_{N^1}$ that reproduce the observed baryon asymmetry. The green, blue, black and red points correspond to $m_{lightest}/\text{eV} = 10^{-2},10^{-2},10^{-2}$ and $10^{-8}$. 
It shows that heavier $m_{{lightest}}$ lead to stronger washout, characterized by $K \gg 1$ (or equivalently, $\kappa \ll 1$). 
This is consistent with the higher minimum value of $m_{N^1}$ observed in the left panel for $m_{lightest} \gtrsim 10^{-3}$ eV.

The corresponding results for the IH is shown in Fig.~\ref{fig:SM3thIH}.  
The left panel shows the minimum values of $m_{N^1}$ obtained for various $m_{lightest}$.  
As in NH case, the minimum value of $m_{N^1}$ becomes independent of $m_{lightest}$ for $m_{lightest} \lesssim 10^{-3}$ eV with the smallest value,  $m_{N^1}^{min} = 6.52\times 10^8$ GeV.
In the right panel, we show $K = \Gamma_1/H$ for various $m_{N^1}$ that reproduce the observed baryon asymmetry. The green, blue, black and red points correspond to $m_{lightest}/\text{eV} = 10^{-2},10^{-2},10^{-2}$ and $10^{-8}$. As in the NH case, larger $m_{{lightest}}$ leads to stronger washout, requiring higher $m_{N^1}^{\text{min}}$.

%%%%%%%%%%%%%%%%%%%%%%%%%%%%%%%%%%%%%%%
\subsection{Non-Thermal Leptogenesis}
%%%%%%%%%%%%%%%%%%%%%%%%%%%%%%%%%%%%%%%
In non-thermal leptogenesis, the RHNs are never thermalized. 
We consider a simple scenario where the lightest RHN $N_R^1$ is produced non-thermally from the decay of SM singlet scalar field $\phi$. 
We may simply identify $\phi$ as the inflaton, but in general, $\phi$ can be any scalar field whose energy density dominates the energy density of the universe. 
We also assume that $\phi$ decay to $N_{2,3}$ is kinematically forbidden, hence, $m_{N^1} < m_\phi/2$ and $ m_\phi < m_{N^{2,3}}$, which is consistent with the assumption of hierarchical RHN masses.

Besides $N_R^1$, $\phi$ can also decay into a pair of SM Higgs doublets ($H$), and their energy densities are determined by the branching ratio ($BR$) of $\phi$, 
\bea
\rho_H &=& \rho_\phi \times BR(\phi \to H^\dagger H), \nonumber\\
\rho_{1} &=& \rho_\phi \times BR(\phi \to N_R^1 N_R^1),  
\eea
where $\rho_\phi$ is the energy density of $\phi$ at the time of its decay.

The resultant lepton asymmetry depends on the reheating temperature, which is determined by details of $\phi$, $N_R^1$, and $H$ decay. 
We expect that the thermal plasma of SM particles is created immediately after $H$ is produced. 
The decay of $N_R^1$ to SM particles can be either prompt or delayed. 
In the following, we consider various possibilities for $N_R^1$ decay: 
\begin{enumerate}[(I)]
\item $N_R^1$ decays promptly after being produced. 
\item $N_R^1$ decays after a while after being produced. 
\end{enumerate}
For each scenario, there are two additional possibilities  
\begin{enumerate}[\; \; (a)]
\item $N_R^1$ is non-relativistic when it decays 
\item $N_R^1$ is relativistic when it decays ($m_{N^1} \ll m_\phi/2$)
\end{enumerate}

\noindent {\bf Scenario $\rm I$-$\rm (a)$ and $\rm I$-$\rm (b)$:}  
Since $N_R^1$ promptly decays after being produced from $\phi$ decay, the energy density of radiation $\rho_{rad}$ and reheat temperature $T_R$ is determined by the energy density of $\phi$ at the time of its decay independently of whether $N_R^1$ is relativistic or non-relativistic. 
Hence, $T_R$ is given by
\bea
\rho_{rad} = \rho_\phi=  \frac{\pi^2}{30} g_*^{SM} T_{R}^4 ,  
\eea
where $g^{SM}_* = 106.75$ is the d.o.f of the SM plasma. 
Since $\phi$ decay to a pair of $N_R^1$ with branching ratio $BR(\phi \to N_R^1 N_R^1)$, the number density of $N_R^1$ is given by
\bea
n_{1} = 2\times \frac{\rho_\phi }{m_{\phi}} \times  BR(\phi \to N_R^1 N_R^1), 
\eea
where $\rho_\phi /m_{\phi}$ is the number density of $\phi$. 
Using $n_L = \epsilon_1 \times n_{1}$ for the lepton asymmetry generated from $N_R^1$ decay, we obtain 
\bea
\frac{n_{L}}{s} = \frac{3}{2} \; \left(\frac{T_R}{m_\phi}\right)\times \epsilon_1 \times BR(\phi \to N_R^1 N_R^1), 
\label{eq:nLs}
\eea
where the entropy density of SM plasma $s$ is given by   
\bea
s &=& \frac{4}{3} \frac{\rho_\phi}{T_R} = \frac{2\pi^2}{45} g_*^{SM} T_{R}^3. 
\eea
The sphaleron process converts the lepton asymmetry to the baryon asymmetry $n_B/s = -C_{sph} \times n_L/s$, where $C_{sph} = 28/79$. Requiring $\frac{n_B}{s} \simeq  8.7 \times  10^{-11}$ to reproduce the observed baryon asymmetry, we obtain 
\bea
\epsilon_1 = -3.27 \times 10^{-10} \frac{1}{\left(2 \times\frac{ T_R}{m_\phi}\right)\times  BR(\phi \to N_R^1 N_R^1)},
\label{eq:eps1nth}
\eea

In non-thermal leptogenesis, the produced $N_{R}^1$ is already out-of-equilibrium and hence $\kappa=1$. To ensure this situation, we simply set $m_{N^1} < T_R$. Considering this condition, we can see that $|\epsilon_1|$ in Eq.~(\ref{eq:eps1nth}) can be minimized to be 
\bea
|\epsilon_1|_{min} \simeq 3.27 \times 10^{-10} 
\label{eq:epsmin}
\eea
for $T_R \simeq m_{N^1} \simeq m_\phi/2$ and $BR_{max}(\phi \to N_R^1 N_R^1) = 1$. \\

\noindent {\bf Scenario $\rm II$-$\rm (a)$:} For simplicity, let us assume that $N_R^1$ is sufficiently long-lived to dominate the energy density of the universe. 
Hence, we can ignore the existing SM thermal plasma. 
$N_R^1$ decay creates the SM thermal plasma with energy density 
\bea
\rho_{rad} = \rho_{1}  = \frac{\pi^2}{30} g_*^{SM} T_{R}^4  ,  
\eea
where $\rho_1$ is the energy density of $N_{R^1}$ at the time of its decay. 
Since $N_R^1$ is non-relativistic, its energy density is given by $\rho_{1} = m_{N^1} n_{1}$. Using $n_L = \epsilon_1 \times  n_{1}$ and $s = \frac{2\pi^2}{45} g_*^{SM} T_R^3$, we obtain 
\bea
\frac{n_{L}}{s} &=& \frac{3}{4} \times \epsilon_1\; \times  \frac{T_R}{m_{N^1}} \; 
\eea
Employing the consistency condition $T_R < m_{N^1}$, 
\bea
\frac{n_{L}}{s} &<& \frac{3}{4} \times \epsilon_1
\eea
leads to the minimum value of $|\epsilon_1|$, which coincides with Eq.~(\ref{eq:epsmin}). \\

\noindent {\bf Scenario $\rm II$-$\rm (b)$:} 
Let $a_\phi = a(t=t_\phi)$ and $a_N= a(t=t_N)$ be the scale factor of the universe at the time of $\phi$ and $N_R^1$ decay, respectively. 
Taking into the expansion of the universe between $t_\phi$ and $t_N$ we obtain 
\bea
\rho_{rad} (t_{N}) &=& \rho_\phi \times BR(\phi \to H^\dagger H) \times \left(\frac{a_\phi}{a_N}\right)^4, \nonumber \\
\rho_{1} (t_{N}) &=& \rho_\phi \times BR(\phi \to N_R^1 N_R^1) \times \left(\frac{a_\phi}{a_N}\right)^4,  \nonumber \\
n_{1} (t_{N}) &=& 2\times \frac{\rho_\phi }{m_{\phi}} \times  BR(\phi \to N_R^1 N_R^1)\times \left(\frac{a_\phi}{a_N}\right)^3, 
\eea
where $\rho_\phi /m_{\phi}$ is the number density of $\phi$. 

Let $T_{SM}$ be the temperature of the SM plasma after $N_R^1$ decay which can be evaluated as 
\bea
\frac{\pi^2}{30} g_*^{SM} (T_{SM})^4 = \rho_{H} (t_N) \; + \; \rho_{1} (t_N) = \rho_{\phi} \times \left(\frac{a_\phi}{a_{N}}\right)^4, 
\label{eq:Tprime}
\eea
where we have used that the sum of branching ratios of $\phi$ is one. 
Using $n_L = \epsilon_1 \times n_{1}$ with the entropy density $s = (2 \pi^2 / 45) g_*^{SM}(T_{SM})^3$, we obtain 
\bea
\frac{n_{L}}{s} &=& \frac{3}{2}\times \epsilon_1
\times \left(\frac{T_{SM}}{m_{\phi}}\right)  \times \left(\frac{a_{N}}{a_\phi}\right) \times BR(\phi \to N_R^1 N_R^1).
\label{eq:nLsrel}
\eea
Note that we have assumed that $N_R^1$ is relativistic when it decays. This means $| {\vec p}_{N^1}| > m_{N^1}$, where $| {\vec p}_{N^1}|$ is the momentum of ${N_R^1}$. We can estimate $| {\vec p}_{N^1}| \simeq \frac{1}{2} m_\phi \left(\frac{a_{\phi}}{a_N}\right) > m_{N^1}$. 
Using the condition $T_{SM} < m_{N^1} \ll m_\phi$ in Eq.~(\ref{eq:nLsrel}), we find 
\bea
\frac{n_{L}}{s} &<& \frac{3}{2}\times \epsilon_1
\times \left(\frac{m_{N^1}}{m_{\phi}}\right)  \times \left(\frac{m_\phi}{2 m_{N^1}}\right) \times BR(\phi \to N_R^1 N_R^1).
\eea
Again, note that this condition leads to the minimum value of $\epsilon_1$, which coincides with Eq.~(\ref{eq:epsmin}).

Since $\epsilon_1 \propto m_{N^1}$ and our focus is on determining the minimum viable value of $\epsilon_1$, we will fix $|\epsilon_1| = 3.27 \times 10^{-10}$ in the following analysis which is the minimum value required for successful non-thermal leptogenesis. 
We now present out results for non-thermal leptogenesis in scenarios with two and three generations of RHNs.

%%%%%%%%%%%%%%%%%%%%%%%%%%%%%%%%%%%%%%%
\subsubsection{Results: SM + 2 RHNs}
%%%%%%%%%%%%%%%%%%%%%%%%%%%%%%%%%%%%%%%

\begin{figure}[t!]
    \centering
\includegraphics[width=0.35\linewidth]{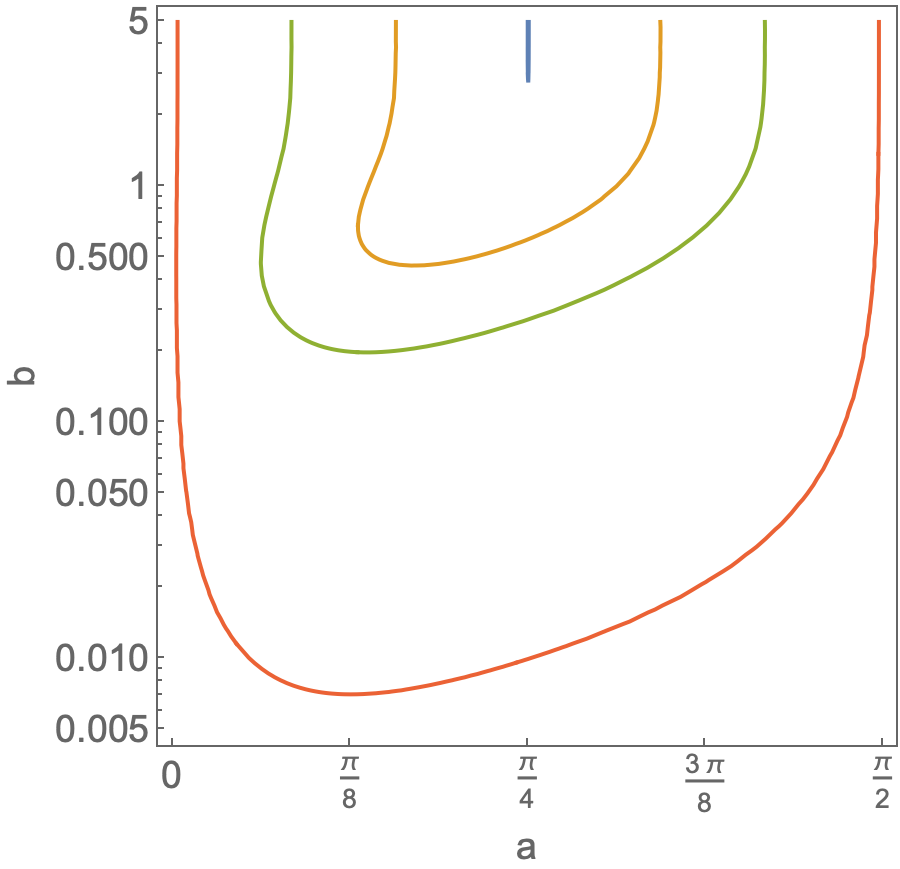}\qquad  \qquad \qquad \qquad 
\includegraphics[width=0.35\linewidth]{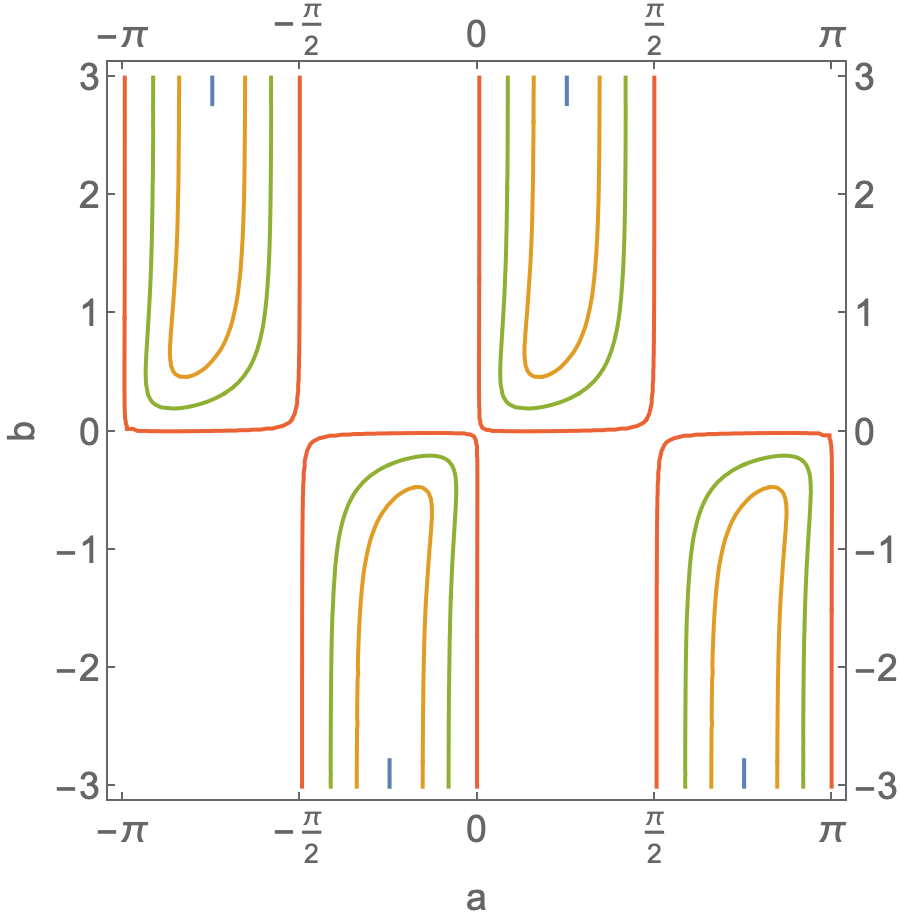}
    \caption{Results for the NH. Both panels show contours of fixed $m_{N^1}$ that yield the observed baryon asymmetry, with the innermost corresponding to $m_{N^1}^{{min}} \simeq 4.00 \times 10^{6}$ GeV, and the outer contours representing $1.2$, $2$, and $50$ times this value.}
    \label{fig:SM2nthNH}
\end{figure}

\begin{figure}[t!]
    \centering
\includegraphics[width=0.35\linewidth]{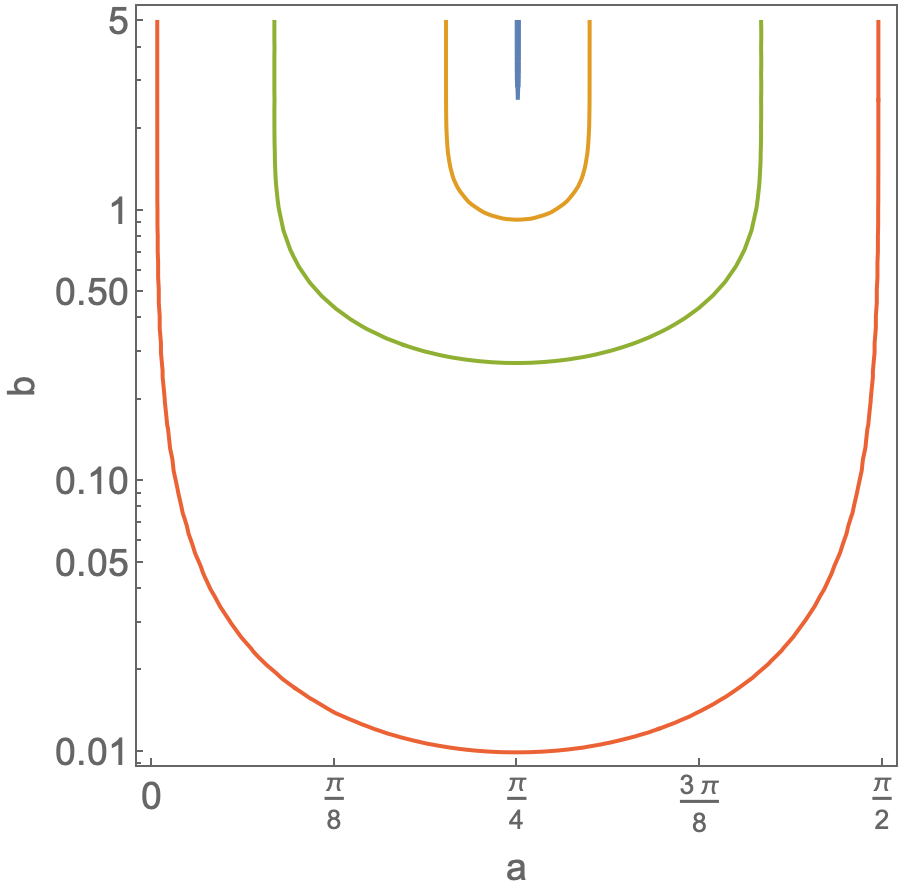}\qquad \qquad \qquad \qquad 
\includegraphics[width=0.35\linewidth]{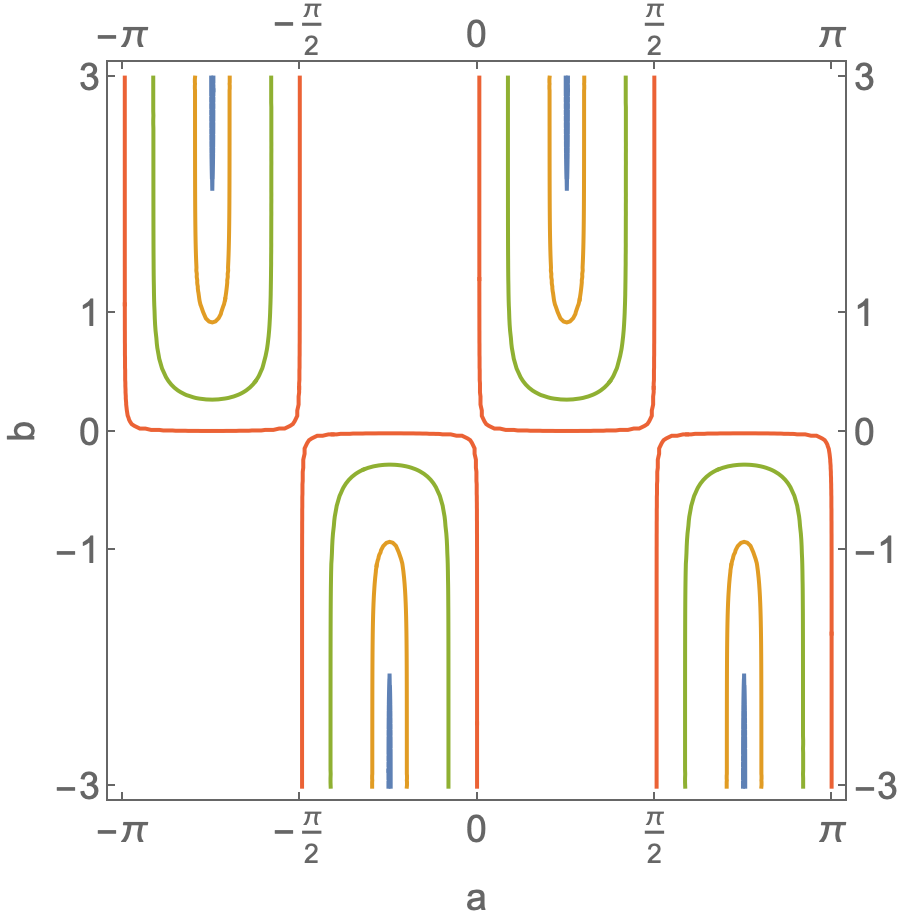}
    \caption{Results for the IH. Both panels show contours of fixed $m_{N^1}$ that yield the observed baryon asymmetry, with the innermost corresponding to $m_{N^1}^{{min}} \simeq 2.22 \times 10^{8}$ GeV, and the outer contours representing $1.05$, $2$,and $50$ times this value.}
    \label{fig:SM2nthIH}
\end{figure}
The explicit formulas for $\epsilon_1$ for the NH and IH are given in Eqs.~(\ref{eq:SM2epsNH}) and (\ref{eq:SM2epsIH}), respectively. 
They are determined by $a$, $b$ and $m_{N^1}$. 
In Fig.~\ref{fig:SM2nthNH}, we show the result for the NH. The left panel shows contours of fixed $m_{N^1}$ in the $a$–$b$ plane to yield $|\epsilon_1|_{min}$ in Eq.~(\ref{eq:epsmin}). The innermost contour corresponding to the minimum value $m_{N^1}^{min} \simeq 4.00 \times 10^{6}$ GeV, with outer contours representing $1.2$, $2$, and $50$ times this minimum. 
The right panel shows that the result for a wider range of $a$ and $b$ with contours corresponding to same benchmark $m_{N^1}$ values. 

In Fig.~\ref{fig:SM2nthIH}, we show the result for the IH. The left panel shows contours of fixed $m_{N^1}$ in the $a$–$b$ plane to yield $|\epsilon_1|_{min}$ in Eq.~(\ref{eq:epsmin}). 
The innermost contour corresponds to the minimum value $m_{N^1}^{min} \simeq 2.22 \times 10^{8}$ GeV, with outer contours representing $1.05$, $2$, and $50$ times this minimum. 
The right panel shows that the result for a wider range of $a$ and $b$ with contours corresponding to same benchmark $m_{N^1}$ values. 

From the analytic expression for $\epsilon_1$, Eq.~(\ref{eq:SM2epsNH}) for NH and Eq.~(\ref{eq:SM2epsIH}) for IH, 
we find that $\epsilon_1$ is maximized for both NH and IH when $a = \frac{\pi}{4} \pm n\pi$ for fixed $b$. The maximum value is independent of $b$ for $b\gg 1$.  
For these parameters, $\epsilon_1$ formula is simplified to 
\bea
|\epsilon_{1}^{NH}| &\simeq& \left(\frac{3 m_{N^1}^{min}}{8\pi v_h^2}\right) \left(m_3 -m_2\right), 
%%%%%%%%%%%%%%%%%%%
\nonumber \\
%%%%%%%%%%%%%%%%%%
|\epsilon_{1}^{IH}| &\simeq&  \left(\frac{3 m_{N^1}^{min}}{8\pi v_h^2}\right)\left(m_2 -m_1\right),   
\label{eq:2genepsmax}
\eea
where $m_2 = 8.71\times 10^{-3}$ eV and $m_3 = 5.00\times 10^{-2}$ eV for the NH and $m_1 = 4.85\times 10^{-2}$ eV  and $m_2 = 4.92\times 10^{-2}$ eV for IH.

%%%%%%%%%%%%%%%%%%%%%%%%%%%%%%%%%%%%%%%
\subsubsection{Result: SM + 3 RHNs}
\begin{figure}[t!]
    \centering
\includegraphics[width=0.47\linewidth]{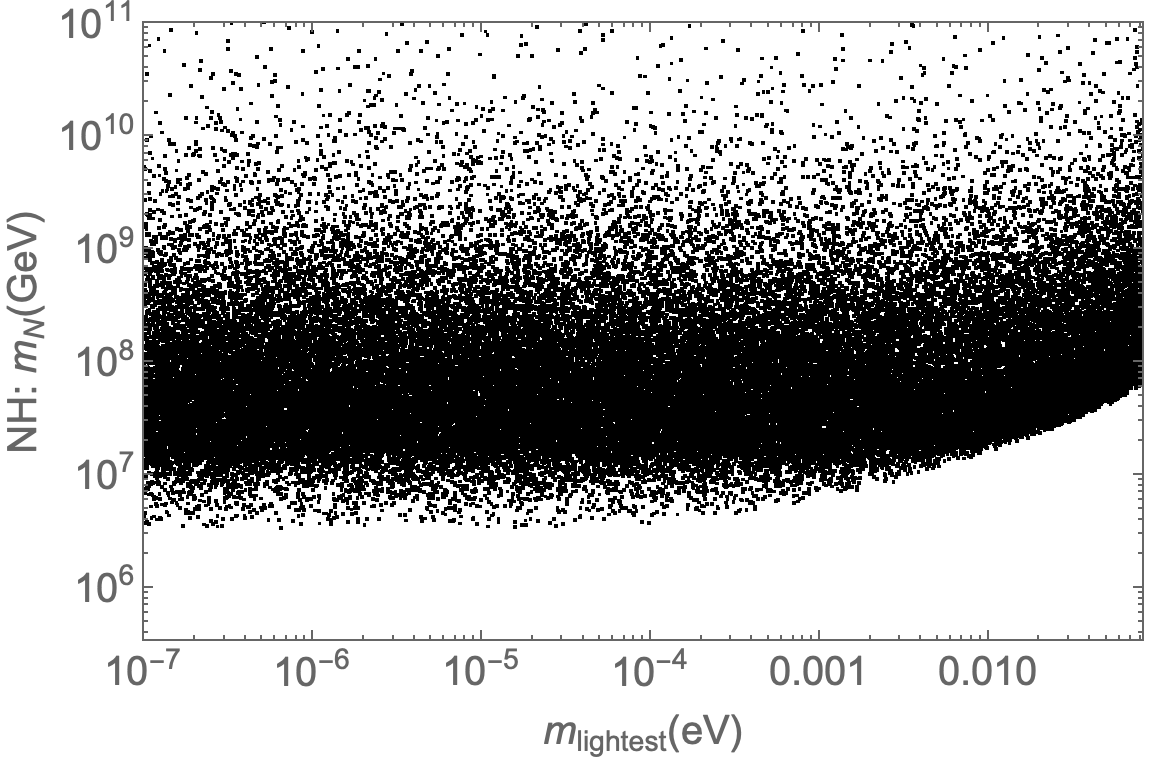}\qquad 
\includegraphics[width=0.45\linewidth]{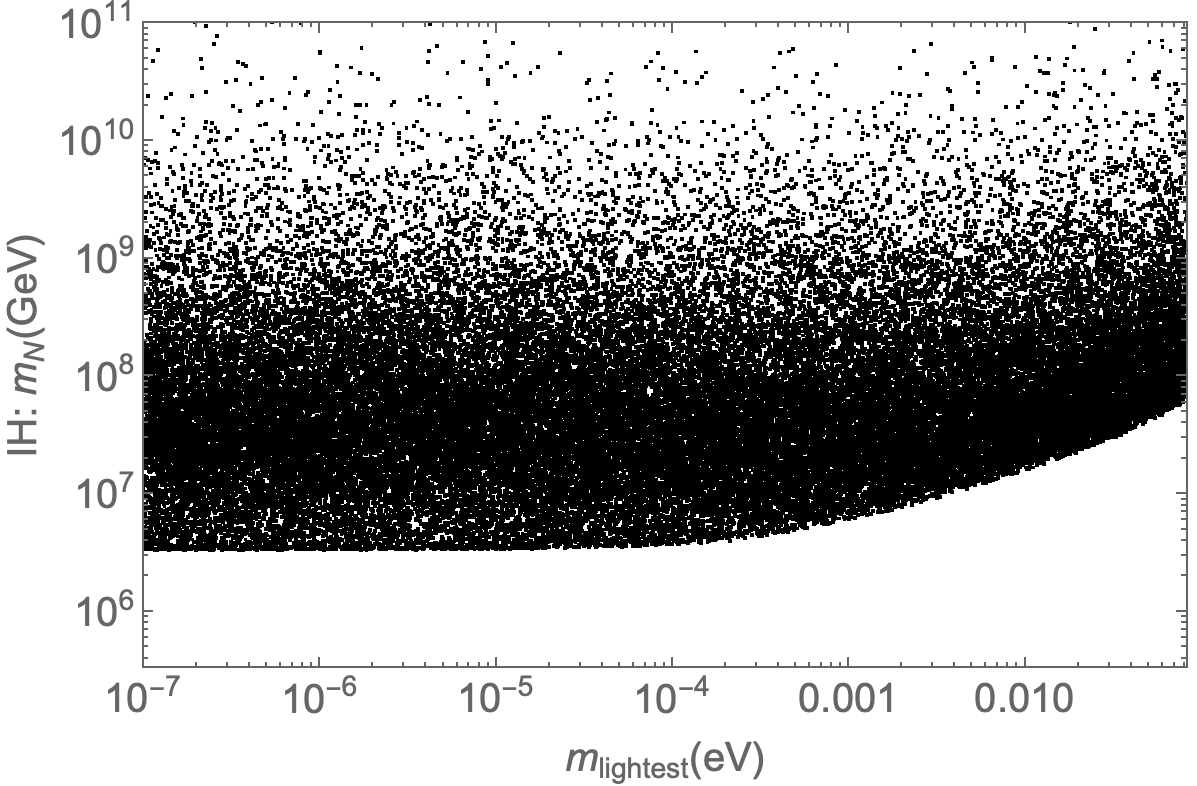}
    \caption{The two plots shows results from a parameter scan. For the NH (IH), the left (right) panel shows shows the minimum values of $m_{N^1}$ obtained for various values of $m_{lightest}$. For NH (IH), we find the smallest value $m_{N^1}^{min} = 3.35 \; (3.32) \times 10^6$ GeV. }
    \label{fig:SM3nthNH}
\end{figure}

%%%%%%%%%%%%%%%%%%%%%%%%%%%%%%%%%%%%%%%
For three generations of RHNs, $\epsilon_1$ is given by Eq.~(\ref{eq:SM3eps}). 
It is determined by $m_{lightest}$, $m_{N^1}$ and the real parameters $a_{12}$, $a_{13}$, $b_{12}$, and $b_{13}$. 
Using the analytical formula, we perform a parameter scan over all the parameters to evaluate the $m_{N^1}^{min}$. 
We show our result in Fig.~\ref{fig:SM3nthNH}. 
For both NH (left panel) and IH (right panel) the plots show the minimum values of $m_{N^1}$ obtained for various $m_{lightest}$ which reproduce $|\epsilon_1|_{min}$ in Eq.~(\ref{eq:epsmin}). 
For both cases, we find that the minimum value of $m_{N^1}$ becomes independent of $m_{lightest}$ for $m_{lightest} \lesssim 10^{-4}$ eV. 
For NH (IH), we find the smallest value $m_{N^1}^{min} = 3.35 \; (3.32) \times 10^6$ GeV.

%%%%%%%%%%%%%%%%%%%%%%%%%%%%%%%%%%%%%%%%%%%%%%%%%%%%%%%%%%%
\section{Type-I Seesaw with Neutrinophillic Higgs Doublet}
%%%%%%%%%%%%%%%%%%%%%%%%%%%%%%%%%%%%%%%%%%%%%%%%%%%%%%%%%%%
The type-I seesaw can also be realized in a two Higgs doublet model, where only the new Higgs doublet ($H_\nu$) couples to SM leptons and RHNs  
\bea  
   {\cal L} \supset  \left(-\sum_{i = 1}^{3} \sum_{j} Y_D^{ij} 
   \overline{N_R^i} \; H_\nu^\dagger\;  \ell_L^j   - \frac{1}{2} \sum_{j} m_{N^j}  \overline{N_{R}^{j~C}} N^j_{R} \right) +{\rm h.c.}
   \label{eq:seesaw2}
\eea 
One way to prevent $N_R$ from coupling to the SM Higgs is to assign a global lepton number $+1$ to $H_\nu$, but not to the RHNs. This is the so-called neutrinophilic Higgs doublet model \cite{Ma:2000cc}. 
Another approach is to consider a gauged $U(1)_X$ extension of the SM, with an alternative charge assignment for $N_R$ \cite{Okada:2018tgy}.

All analytic expressions derived for the seesaw mechanism and leptogenesis in the case with the SM Higgs doublet remain unchanged, except that all instances of $v_h$ are replaced by $v_2$, where $\langle H_\nu \rangle = (v_2/\sqrt{2}, 0)^T$. 
To avoid observable lepton flavor violation induced by the new Dirac Yukawa coupling, we require $v_2 \gtrsim 10$ MeV \cite{Guo:2017ybk}.

%%%%%%%%%%%%%%%%%%%%%%%%%%%%%%%%%%%%%%%
\subsubsection{Results: SM + 2 RHNs}
%%%%%%%%%%%%%%%%%%%%%%%%%%%%%%%%%%%%%%%
{\bf Non-thermal Leptogenesis:} Let us first consider the non-thermal case and require $\epsilon_1 = -3.27\times 10^{-10}$. 
The explicit formulas for $\epsilon_1$ for NH and IH are given in Eqs.~(\ref{eq:SM2epsNH}) and (\ref{eq:SM2epsIH}), respectively, with the replacement $v_h \to v_2$. 
Hence, $\epsilon_1$ is determined by four free parameters, $m_{N^1}$, $a$, $b$, and $v_2$. 
Since $n_B/s \propto \epsilon_1$ and $\epsilon_1 \propto m_{N^1}/v_2^2$, the lower bound on $m_{N^1}$ can be obtained by appropriately scaling the bound derived for the SM Higgs case.
Particularly, we can do so by using Eq.~(\ref{eq:2genepsmax}) with the replacement $v_h \to v_2$ and setting $|\epsilon_1| = 3.27\times 10^{-10}$, 
\bea
{\rm NH:}\; m_{N^1}^{min} &\simeq& 4.00 \times 10^6 \; {\rm GeV}\left(\frac{v_2}{246\; {\rm GeV}}\right)^2 
%%%%%%%%%%%%%%%%%%%
\nonumber \\
%%%%%%%%%%%%%%%%%%
{\rm IH:}\; m_{N^1}^{min} &\simeq& 2.22 \times 10^8 \; {\rm GeV}\left(\frac{v_2}{246\; {\rm GeV}}\right)^2 
\eea
Since the sphaleron process which is essential to generate baryon asymmetry from RHN decay become inactive below $132.7$ GeV \cite{DOnofrio:2014rug}, $m_{N^1}^{min} \simeq 132$ GeV for both NH and IH. \\

\begin{figure}[t!]
    \centering
\includegraphics[width=0.47\linewidth]{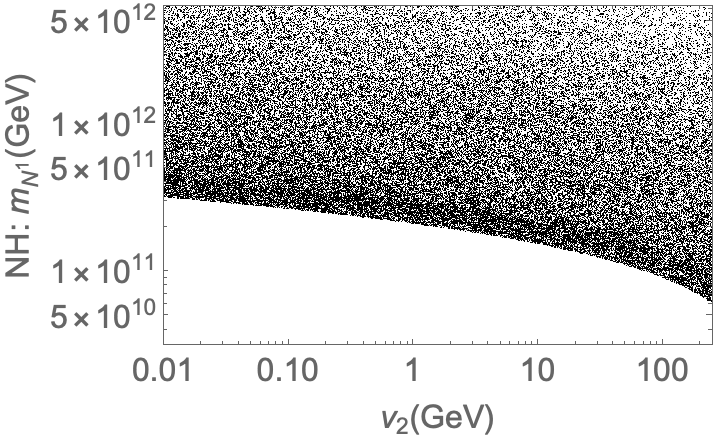}\qquad 
\includegraphics[width=0.47\linewidth]{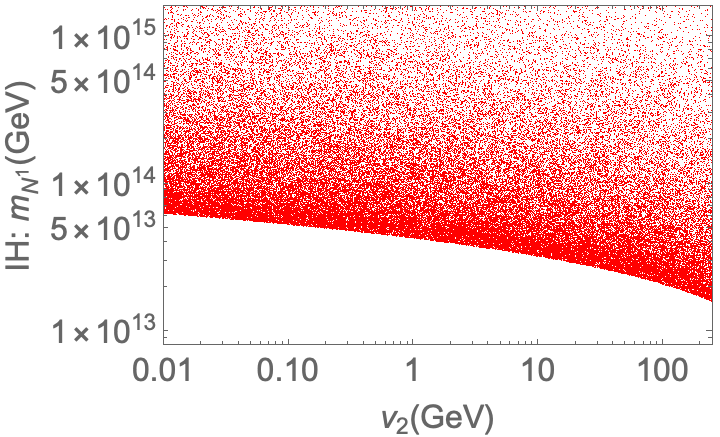}
    \caption{The two plots show the results of a parameter scan for a thermal leptogenesis scenario with three generations of RHNs and Type-I seesaw with a new neutrinophilic Higgs doublet. For NH (left) and IH (right), the plots show the minimum values of $m_{N^1}$ obtained for various values of $v_{2}$, which is the VEV of the new Higgs doublet.}
    \label{fig:2HD2th}
\end{figure}

\noindent{}{\bf Thermal Leptogenesis:} For thermal leptogenesis $n_B/s$ is given by Eq.(\ref{eq:epsthermal}) with the explicit formulas for $\epsilon_1$ and $\Gamma_1$ provided in Eqs.(\ref{eq:SM2epsNH}) and (\ref{eq:SM2epsIH}) for NH and IH, respectively, with the replacement $v_h \to v_2$. 
Hence, $n_B/s$ is determined by four free parameters, $m_{N^1}$, $a$, $b$, and  $v_2$. 
We perform a parameter scan over all the parameters to evaluate the $m_{N^1}^{min}$. 
We show our results in Fig.~\ref{fig:2HD2th}, where the minimum values of $m_{N^1}$ are plotted for various $v_2$, with the NH (IH)  shown in the left (right) panel. 
%For both NH and IH, we find that the lowest allowed value of $v_2 = 10$ MeV leads to the largest minimum value of $m_{N^1}$ with $m_{N^1} = 3.25 \times 10^{10}$ GeV for NH and $6.4 \times 10^{13}$ GeV for IH. 
The increase in $m_{N^1}^{\text{min}}$ for small values of $v_2$ is necessary to compensate for the enhanced washout. This behavior can be qualitatively understood by examining the $v_2$ dependence of $\epsilon_1$ and $\kappa$ in Eq.~(\ref{eq:dilution}).

%%%%%%%%%%%%%%%%%%%%%%%%%%%%%%%%%%%%%%%
\subsubsection{Results: SM + 3 RHNs}
%%%%%%%%%%%%%%%%%%%%%%%%%%%%%%%%%%%%%%%
\begin{figure}[th!]
    \centering
\includegraphics[width=0.46\linewidth]{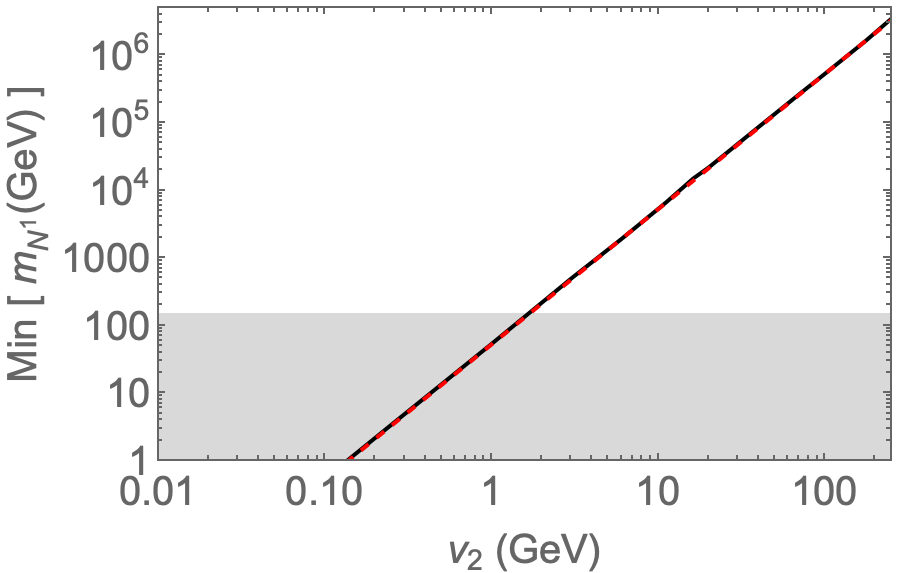}\qquad \; \; 
\includegraphics[width=0.46\linewidth]{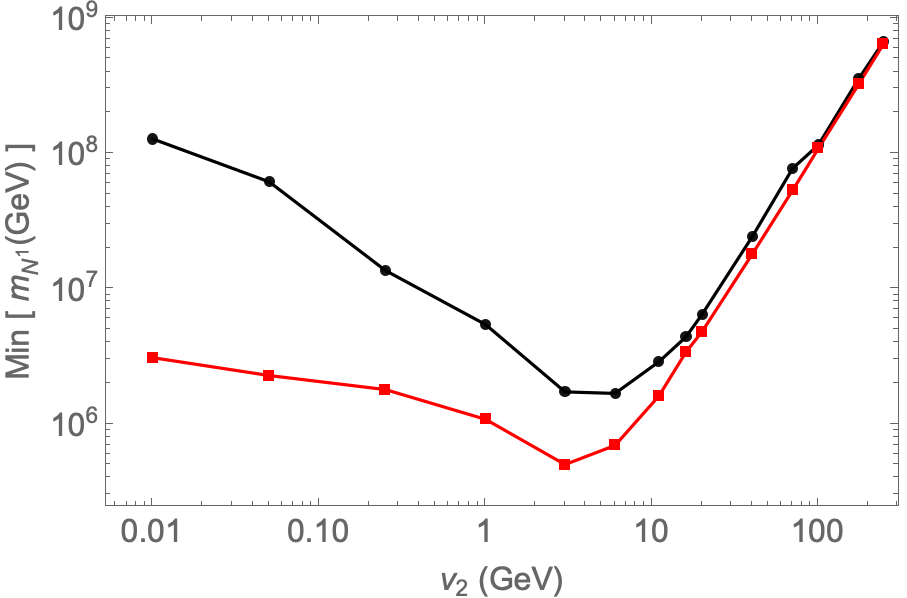}\qquad 
    \caption{The two plots show the results of a parameter scan for leptogenesis scenario with three generations in the neutrinophilic Higgs doublet case.
    The left panel shows the minimum values of $m_{N^1}$ obtained for the non-thermal leptogenesis scenario as a function of $v_{2}$. For both NH (black) and IH (dashed red), values with $m_{N^1}^{\text{min}} < 132$ GeV is excluded due to the absence of sphaleron processes, as indicated by the gray shaded region.
    The right panel shows the result for the minimum value of $m_{N^1}$ as a function of $v_2$ in thermal leptogenesis. For NH (red) and IH (black), we find that $m_{N^1}^{{min}} = 1.72 \times 10^{6}$ GeV and $5.11 \times 10^{5}$ GeV, respectively. 
    }
    \label{fig:2HD3}
\end{figure}

{\bf Non-thermal Leptogenesis:} For three generations of RHNs, $\epsilon_1$ is given by Eq.~(\ref{eq:SM3eps}), with the replacement $v_h \to v_2$. 
Hence, $\epsilon_1$ is determined by six free parameters, $m_{N^1}$, $m_{\text{lightest}}$, $a_{12}$, $a_{13}$, $b_{12}$, $b_{13}$, and $v_2$. 
We perform a parameter scan over all the parameters to evaluate the $m_{N^1}^{min}$. 
We show our results in the left panel of Fig.~\ref{fig:2HD3}, where the minimum values of $m_{N^1}$ are plotted for various $v_2$. 
The lines for NH (solid black) and IH (dashed red) overlap each other. 
The gray shaded region corresponding to $m_{N^1} \lesssim 132$ GeV is excluded because sphaleron process needed to to generate baryon asymmetry from RHN decay become inactive below 131.7 GeV \cite{DOnofrio:2014rug}. \\

\noindent{}{\bf Thermal Leptogenesis:} For three generations, $n_B/s$ is given by Eq.(\ref{eq:epsthermal}) with $\epsilon_1$ and $\Gamma$ given by Eq.~(\ref{eq:SM3eps}) with $v_h$ substituted for $v_2$. 
Hence, $n_B/s$  is determined by six free parameters, $m_{N^1}$, $m_{\text{lightest}}$, $a_{12}$, $a_{13}$, $b_{12}$, $b_{13}$, and $v_2$. 
We perform a parameter scan over all the parameters to evaluate the $m_{N^1}^{min}$. 
We show our results in the right right panel of Fig.~\ref{fig:2HD3}, where the minimum values of $m_{N^1}$ are plotted for various $v_2$. 
The solid black (red) lines are the results for NH (IH). 
We find that $m_{N^1}^{{min}} = 1.72 \times 10^{6}$ GeV for NH and $5.11 \times 10^{5}$ GeV for IH, both of which are realized for $v_2$ values of a few GeV. 
As discussed in the case of the scenario with 2 RHNs, larger $m_{N_1}^{{min}}$ for small values of $v_2$ is necessary to compensate for the enhanced washout. 
For $v_2 \gg 5$ GeV, we find that the washout is highly suppressed with $\kappa = 1$. 
In this case $n_B/s \propto \epsilon_1 \propto m_{N^1}/v_2^2$, which qualitatively explains the behavior of the curves for $v_2 \gg 5$ GeV.

%%%%%%%%%%%%%%%%%%%%%%%%%%%%%%%%%%%%%%%
\section{Summary}
%%%%%%%%%%%%%%%%%%%%%%%%%%%%%%%%%%%%%%%
\begin{table}[th!]
\centering
\begin{tabular}{|c|c|c|c|c|}
\hline
 & \multicolumn{2}{c|}{2 generations} & \multicolumn{2}{c|}{3 generations} \\
\cline{2-5}
 & Normal & Inverted & Normal & Inverted \\
\hline
Thermal & $6.24 \times 10^{10}$ GeV & $1.62 \times 10^{13}$ GeV & $6.68 \times 10^{8}$ GeV & $6.52 \times 10^{8}$ GeV\\
\hline
Non-Thermal & $4.00 \times 10^{6}$ GeV & $2.22 \times 10^{8}$ GeV & $3.35 \times 10^{6}$ GeV & $3.32 \times 10^{6}$ GeV\\
\hline 
\end{tabular}
\caption{Parameter scan results for the lowest mass of the lightest RHN in the conventional Type-I seesaw scenario.}
\label{tab:1}
\end{table}

\begin{table}[th!]
\centering
\begin{tabular}{|c|c|c|c|c|}
\hline
 & \multicolumn{2}{c|}{2 generations} & \multicolumn{2}{c|}{3 generations} \\
\cline{2-5}
 & Normal & Inverted & Normal & Inverted \\
\hline
Thermal & $6.24 \times 10^{10}$ GeV & $1.62 \times 10^{13}$ GeV & $1.72 \times 10^{6}$ GeV & $5.11 \times 10^{5}$ GeV\\
\hline
Non-Thermal & $132$ GeV & $132$ GeV & $132$ GeV & $132$ GeV\\
\hline
\end{tabular}
\caption{
Parameter scan results for the lowest mass of the lightest RHN in the neutrinophillic Higgs doublet Model for the doublet VEV in the range $10^{-2} <v_2/{\rm GeV}< 246$. }
\label{tab:2}
\end{table}

The observed neutrino oscillations and baryon asymmetry, unexplained by the SM, can both be accounted for by extending the SM to include Majorana RHNs. 
Tiny neutrino masses naturally arise in the Type-I seesaw mechanism. The RHNs can also generate the baryon asymmetry via leptogenesis, in which, the lepton asymmetry is generated by the out-of-equilibrium decay of RHNs and then it is converted to a baryon asymmetry via the sphaleron process.

The Dirac Yukawa couplings play the crucial role for both Type-I seesaw and leptogenesis. 
In order to reproduce the neutrino oscillation data, we employ the general parameterization for the Dirac Yukawa couplings by introducing a complex orthogonal matrix. 
Using the general parameterization, we derived an analytic expression for the CP asymmetry parameter in leptogenesis.

Given that at least two RHNs are required to reproduce all neutrino oscillation data, we consider cases with two and three generations of RHNs. For each case we consider both thermal and non-thermal leptogenesis scenarios. 
We focused on a hierarchical RHN mass spectrum, where the lightest RHN determines the lepton asymmetry. 
Using the analytic formula for the CP asymmetric parameter, we performed parameter scan to find the lowest mass of the lightest RHN necessary for reproducing the observed baryon asymmetry. 

%In thermal leptogenesis, RHNs are in thermal equilibrium with the SM plasma in the early universe, whereas in non-thermal leptogenesis, RHNs are never thermalized.  
%Additionally, in thermal leptogenesis, the lepton asymmetry produced by RHN decay can be washed out by inverse processes active during RHN decoupling.  

In Table \ref{tab:1} we summarize our result for the lowest mass of the lightest RHN, $m_{N^1}^{min}$, in the conventional Type-I seesaw scenario. 
For both thermal and non-thermal scenarios with two generations of RHNs, we find that $m_{N^1}^{min}$ is lower for the NH than for the IH. 
In the case with three generations, we find that $m_{N^1}^{min}$ is almost the same for both cases.

We also examined Type-I seesaw generated by the RHN coupling with SM lepton doublet and a new neutrinophillic Higgs doublet. The vacuum expectation value of the new Higgs doublet $v_2$ is also a free parameter.  
We summarize our results in Table \ref{tab:2} for $10^{-2} <v_2/{\rm GeV}< 246$. 
In non-thermal leptogenesis, for both two and three RHN generations, we found $m_{N^1}^{min} \simeq 132$ GeV which is the sphaleron decoupling temperature. 
For thermal leptogenesis with two generations of RHN, $m_{N^1}^{min}$ listed in the table correspond to $v_2 = 246$ GeV. 
For the case with three generations, $m_{N^1}^{min}$ occurs for $v_2 \simeq 3$ GeV. 

\vspace{15pt}

\noindent {\it Note added}: During the completion of this work, we became aware of Ref.~\cite{Granelli:2025cho}, which contains some overlap with the present study.

%%%%%%%%%%%%%%%%%%%%%%%%%%%%%%%%%%
\section*{Acknowledgments}
%%%%%%%%%%%%%%%%%%%%%%%%%%%%%%%%%%
The work of N.O. is supported in part by the United States Department of Energy Grants DE-SC0012447 and DC-SC0023713.

\printbibliography

\end{document}